\pdfoutput=1

\documentclass[num-refs]{nbdt-article}

\usepackage{siunitx}

\papertype{Original Article}

\title{Modelling Spontaneous Firing Activity of the Motor Cortex in a Spiking Neural Network with Random and Local Connectivity}

\author[1]{Lysea Haggie}
\author[1]{Thor Besier}
\author[1, 2]{Angus McMorland}

\affil[1]{Auckland Bioengineering Institute, University of Auckland, Auckland, New Zealand}
\affil[2]{Department of Exercise Sciences, University of Auckland, Auckland, New Zealand}

\corraddress{Lysea Haggie, Level 8 Auckland Bioengineering Institute, The University of Auckland, Auckland, 1010, New Zealanad}
\corremail{lysea.haggie@auckland.ac.nz}

\fundinginfo{L.H. was supported by funding from Callaghan Innovation. The funders had no role in the study design or decision to publish.}

\runningauthor{Haggie et al.}

\begin{document}

\maketitle

\begin{abstract}

Computational models of cortical activity can provide insight into the mechanisms of higher-order processing in the human brain including planning, perception and the control of movement. Activity in the cortex is ongoing even in the absence of sensory input or discernible movements and is thought to be linked to the topology of the underlying cortical circuitry \citep{Ringach2009}. However, the connectivity and its functional role in the generation of spatio-temporal firing patterns and cortical computations are still unknown.

Movement of the body is a key function of the brain, with the motor cortex the main cortical area implicated in the generation of movement. We built a spiking neural network model of the motor cortex which incorporates a laminar structure and circuitry based on a previous cortical model by \citet{Potjans2014}. A local connectivity scheme was implemented to introduce more physiological plausibility to the cortex model, and the effect on the rates, distributions and irregularity of neuronal firing was compared to the original random connectivity method and experimental data. Local connectivity increased the distribution of and overall rate of neuronal firing. It also resulted in the irregularity of firing being more similar to those observed in experimental measurements, and a reduction in the variability in power spectrum measures. 

The larger variability in dynamical behaviour of the local connectivity model suggests that the topological structure of the connections in neuronal population plays a significant role in firing patterns during spontaneous activity. This model took steps towards replicating the macroscopic network of the motor cortex, replicating realistic spatiotemporal firing to shed light on information coding in the cortex. Large scale computational models such as this one can capture how structure and function relate to observable neuronal firing behaviour, and investigates the underlying computational mechanisms of the brain.

\end{abstract}
\section{Introduction} \label{intro}
The motor cortex is critical for producing movement. However, neurons in the motor cortex show patterns of ongoing activity even during resting state periods with no discernible movement \citep{Schieber2011}. Baseline spontaneous activity, which is observed in awake and conscious experimental subjects, is differentiable from task-induced states in recordings of neuron spiking behaviour \citep{Dabrowska2021, Velliste2014, Churchland2012, Sauerbrei2020}. Spontaneous activity of the brain at rest has spatially structured patterns of activity and is considered to play an important role in neural encoding for brain processes \citep{Ringach2009, Deco2013}. Previously, activity in the cortex was thought to be driven by external sensory input, but it recently has been suggested that spontaneous activity arises from the dynamics of the neuronal circuitry and inputs modulate and modify the dynamics of the network rather than drive it \citep{Destexhe2011, Chen2019}. Thus, there is an interest in understanding the ongoing behaviour of neural activity, how it reflects the cortical circuitry and shapes the variable functions of the cortex \citep{Tan2015, Destexhe2011, Baker1999}.

Spontaneous activity in the motor cortex, similar to other cortical areas, is characterised by irregular firing of individual neurons, with low average firing rates ($<10$~Hz) but a wide frequency range ($0$ - $100$~Hz) \citep{Dabrowska2021, Tomov2014, Hahn2010, Lacey2014, Borges2020}. Firing rates in cortical networks measured \textit{in vivo} are reported to have highly positively skewed, long-tail distributions \citep{Roxin2011}. Neuron populations in the cortex also display oscillatory firing activity covering a broad frequency spectrum ranging from less than one hertz to hundreds of hertz \citep{Buzsaki2004, Tomov2014}. In the motor cortex, beta oscillations ($13$--$30$ Hz) are consistently observed in electroencephalography (EEG) measurements during resting states or prior to movement, and disappear during movement \citep{Espenhahn2019, Schmidt2019}. 

The origin of the variability in the firing of cortical neurons is unknown and could be due to morphology of dendrites resulting in non-linear integration of inputs, or the properties of the network and synaptic coupling \citep{Softky1993, Renart2010, Ostojic2014, Khanna2017}. The variability in neuronal activity is also believed to play a significant role in information encoding in the cortex \citep{Habenschuss2013, Hawkins2017, Shamir2014}. Recently, the generation of neocortical activity has been addressed through computational models of somatosensory cortical circuits based on anatomical and physiological data, which generate realistic spontaneous dynamics \citep{Markram2015, Potjans2014}. However, specific model structure or features have not been clearly linked to cortical firing patterns. 

The cortex is thought to have a consistent laminar organisation and patterns of connectivity across different regions \citep{Douglas2004, DaCosta2010, Hawkins2017}. A canonical circuit which generally defines recurrent connections in each layer, layer 4 and 6 as input layers, and layers 2/3 and 5 as output layers, was based originally on \citet{Hubel1962}'s work on the cat visual cortex and a similar pattern observed in other cortical areas in primate studies including the auditory cortex, somatosensory cortex and motor cortex \citep{Douglas2004, Ghosh1988, Huntley1991, DaCosta2010}. \citet{Binzegger2004} quantified connectivity of the cortical circuit through \textit{in vivo} intracellular recordings and morphometry of cell types and laminar distribution. This and other photostimulation and optogenetic studies of the mouse motor cortex have shown a dominant connection of layer 2/3 neurons to layer 5 neurons \citep{Weiler2008, Hooks2011, Hooks2013}.

Experiments involving \textit{in vivo} extracellular injections of neuronal tracers in the cortex observe 'patchy' projections, in which synapses are highly localised \citep{Huntley1991} (also see \citet{Voges2010} for review). Recent photostimulation experiments and digital image reconstruction suggest a distance-based connectivity model with strong connectivity within 0.2~mm of the soma and connectivity decreasing as a function of distance \citep{Bender2003, Shepherd2005}. Diffusion tensor imaging also shows small-world, patchy connections in the cortex and greater local connectivity than "random" connections \citep{Gong2009}. Localised connections are also thought to be optimal in regards to minimising wiring but maximising connectivity between nodes in a network. Local connections with some long range connections may be an efficient method of information transfer in the brain \citep{Voges2010}. Structural variations in network topology influence the spatio-temporal activity of the cortex, with changes in connectivity patterns resulting in different dynamics \citep{Senk2022}.

Large-scale neural network models can be used to explore how network structure influences the generation of cortical activity. \citet{Voges2010} modelled the horizontal connectivity in the neocortex and showed complex activity patterns arise in structured, spatially clustered synapses. There recently has also been the development of larger-scale cortical models which contain tens of thousands of spiking neuron models \citep{Esser2005, Potjans2014}. However, previous models of the motor cortex have been limited in replicating physiological detail and mainly been focused on generating movement dynamics or responses to stimulation \citep{Farokhniaee2021, Lee1998, Esser2005, Manola2007}. The role of network connectivity in generating spontaneous cortical activity in physiologically-based spiking neural network models has not been explored. 

Computational models can aid in the understanding of the electrophysiological mechanisms which underlie the generation of spontaneous activity. In this study, a large-scale spiking neural network containing over 38,000 neurons and 150 million synapses, based on previous modelling work by \citet{Potjans2014}, was implemented in the Python-based neural network simulator, Brian2 \citep{Stimberg2019}. The model was developed to replicate the spontaneous firing behaviour of the motor cortex and explore the effect of local connectivity on the network behaviour. The aim of this work was to provide insights into resting state motor cortex dynamics using a spiking neural network model and investigate connectivity as a potential source of variability and efficient information transmission in spontaneous firing behaviour. 
\section{Methods} \label{methods}

This model was based on previous work by \citet{Potjans2014} and implemented in Brian2 with reference to source code from \citet{Shimoura2018}. The original cortical circuit model was adapted to represent a $1~\text{mm}^2$ surface area of the motor cortex. The model follows a laminar structure, grouping the neurons into four layers, 2/3, 4, 5, and 6. Each layer was further divided into excitatory and inhibitory cell groups. Layer 1 was ignored due to its low density of neuronal cell bodies \citep{Muralidhar2014}. Connection weights in the circuit are shown in figure \ref{fig:circuit} below. 

\begin{figure}[h]
\centering
\includegraphics[scale=0.3]{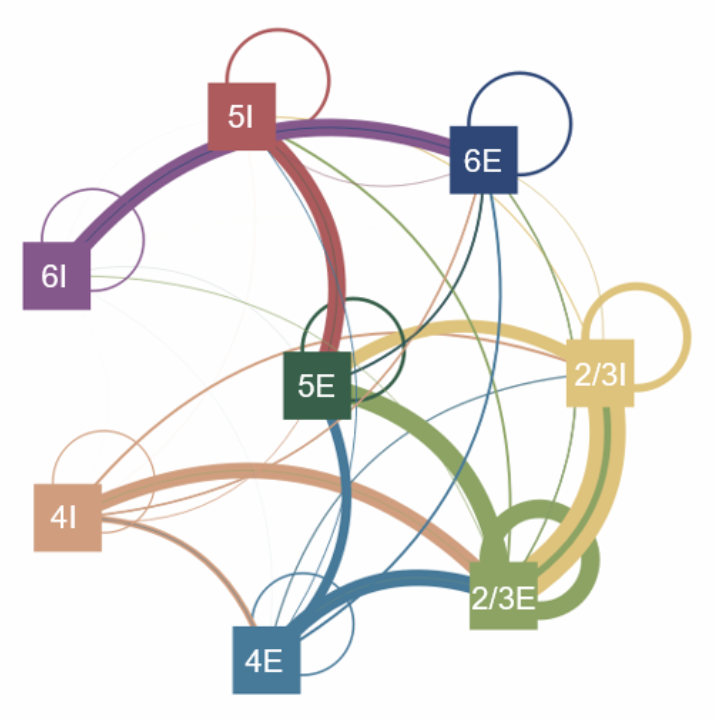}
\caption{Circuit diagram of the motor cortex model showing connections between the populations of excitatory (E) and inhibitory (I) neuron groups. Groups are organised according to their layer. The colour of the lines indicates the source group. Thickness represents the relative number of connections and weighting of synapses.}
\label{fig:circuit}
\end{figure}

Individual neuron dynamics were governed by leaky-integrate-and-fire equations simulated using the linear state updater with a time step of $0.1$~ms. Equation~\ref{eq:dvdt} describes the membrane potential ($V$) of each neuron, where $\tau_{m}$ is the time constant, $C_{m}$ is the membrane capacitance, $V_{r}$ is the reset value for the membrane potential following a spike and $I_{syn}$ is the total input current described by equation~\ref{eq:Isyn}. Action potentials were fired whenever $V(t)$ became more positive than the threshold ($\theta$). On the firing of a presynaptic neuron, the synaptic current of the postsynaptic neuron ($I_{syn}^{post}$) was changed by the value determined by the conductance ($g$) multiplied by the weight ($w$) after a delay ($d$) which accounts for the finite time interval of an action potential propagation in a presynaptic neuron (equation \ref{eq:weight}). The delay between the spiking activity of a presynaptic neuron and postsynaptic neuron was drawn from a normal distribution with a mean of 1.5~ms for excitatory neurons and 0.8~ms for inhibitory neurons; the standard deviation of delay times was half of the mean value. A shorter inhibitory delay was used in the model as inhibitory neurons generally make more local connections over smaller distances \citep{Isaacson2011, Holmgren2003, Yoshimura2005, Packer2011, Fino2013}. In unmyelinated axons, each millimetre could introduce a conduction delay of at least 2~ms \citep{Thomson2002}. Cortical neurons exhibit a range of myelination, though inhibitory interneurons also show more myelination, thus also potentially increasing conduction speed \citep{Call2021}. Our model only incorporated 1~mm of cortical area and so delays were defined as less than 2~ms with shorter delays for inhibitory neurons, and were also independent of layer as in \citet{Potjans2014}. Following a spike being fired, the membrane voltage was reset to -65~mV. The refractory period of the neuron was 2~ms meaning another spike could not be fired within that period of time following a previous spike. As in the original \citet{Potjans2014} model, the conductance value in the layer 4E to layer 2/3E connection was doubled compared to the other excitatory connections. Parameter descriptions and values are given in table \ref{table:parameters}.

\begin{equation}
\frac{dV\left(t\right)} {dt}
= \frac{-(V\left(t\right) - V_{r})}
{\tau_{m}} + \frac{I_{syn}\left(t\right)}{C_{m}}
\label{eq:dvdt}
\end{equation}

\begin{equation}
\frac{dI_{syn}}{dt} = -\frac{I_{syn}}{{\tau_{syn}}}
\label{eq:Isyn}
\end{equation}

\begin{equation}
\label{eq:weight}
\begin{split}
I_{syn}^{post}\left(t + d \right) &= I_{syn}^{post}\left(t + d \right) + g \cdot w \\
&\textrm{when presynaptic neuron V(t)} \geq \theta
\end{split}
\end{equation}

\begin{table}[bt]
\centering
\caption{Parameters for neuron model. Taken from \citet{Potjans2014}.}
\label{table:parameters}
\begin{threeparttable}
\begin{tabular}{lll}
\headrow
\thead{Symbol} & \thead{Name} & \thead{Value}\\
$C_{m}$ & Membrane Capacitance & 250 pF \\
$\theta$ & Threshold & -50 mV \\
$\tau_{ref}$ & Refractory Time & 2 ms\\
$\tau_{m}$ & Membrane Time Constant & 10 ms \\
$\tau_{syn}$ & Synaptic Time Constant & 0.5 ms \\
$V_{r}$ & Reset Value & -65 mV\\
$g_{e}$ & Excitatory Conductance & 1 \\
$g_{i}$ & Inhibitory Conductance & -4 \\
$w$ & Synaptic Weight & 87.8~pA \\
$d_{e}$ & Mean Excitatory Delay & 1.5 ms\\
$d_{i}$ & Mean Inhibitory Delay & 0.8 ms\\
\hline  
\end{tabular}
\end{threeparttable}
\end{table}

\begin{table}[bt]
\centering
\caption{Neuron groups in the motor cortex model, the proportions of neurons in each group (kept constant in sensitivity analysis), and absolute number of neurons in each neuron group in the final model.}
\label{table:num neurons}
\begin{threeparttable}
\begin{tabular}{ccccccccc}
\headrow
\thead{Group} & \thead{2/3E} & \thead{2/3I} & \thead{4E} & \thead{4I} & \thead{5E} & \thead{5I} & \thead{6E} & \thead{6I}\\
\thead{Proportion} & 0.268 & 0.076 & 0.063 & 0.014 & 0.284 & 0.071 & 0.187 & 0.038 \\
\thead{Absolute} & 10332 & 2916 & 2412 & 540 & 10944 & 2736 & 7200 & 1476 \\
\hline  
\end{tabular}
\begin{tablenotes}
\item E, excitatory; I, inhibitory. Number indicates cortical layer.
\end{tablenotes}
\end{threeparttable}
\end{table}

Neurons were distributed over eight populations, according to the numbers in table~\ref{table:num neurons}. Neuron numbers were adjusted from the original \citet{Potjans2014} model, switching the proportion of layer 4 and 5 cells based on the motor cortex having a dominant layer 5 connectivity and sparser cell distributions than other layers of the cortex (ie. the somatosensory and visual cortices on which the original model was based) \citep{Young2013}. Previously, the motor cortex has thought to lack a distinct layer 4 but recent studies support a functional, albeit small, layer 4 in the motor cortex with similar connections to layer 4 in the somatosensory cortex, notably as an input pathway from the thalamus \citep{Yamawaki2014, Barbas2015}. The superficial layers of 2/3 and 4 contain approximately 40\% of the neurons, layer 5 has 35\% of the neurons and layer 6 has 25\% of the neurons, which is similar to reported physiological experimental data \citep{Amunts1995}. The E/I ratio in this model is on average 24\%. With inhibitory neurons making up 28.2\% 24.8\%, 22.4\%, 20.5\% for layers 2/3, 4, 5 and 6, respectively.
The proportion of GABA cells in motor area has been reported as 24.2\% with slightly lower percentage of GABA in deeper layers \citep{Hendry1987}. A sensitivity analysis was carried out to investigate the effect of the total number of neurons in the network, to select a minimal number of neurons at which the rate and irregularity of firing activity converges.

\citet{Potjans2014} original connectivity map integrated the findings of multiple anatomical and electrophysiological studies to estimate the probabilities of connection and number of synapses formed between each neuron group. Combining these data from different species and areas, including rat visual and somatosensory areas and cat visual striate cortex, builds on the theoretical framework suggesting an equivalency across cortical areas. Comparative studies of primary motor cortex also show functional preservation across mammals \citep{Bakken2021}. The connectivity profile was based on a modified version of Peter's rule which proposes that the number of synapses is dependent on the number of neurons collocated in the presynaptic and postsynaptic layers and a probability value of connection between layers, derived from experimental data \citep{Peters1993, Braitenberg1998, Binzegger2004, Rees2017}. 

\begin{figure}[ht]
\centering
\includegraphics[scale=0.5]{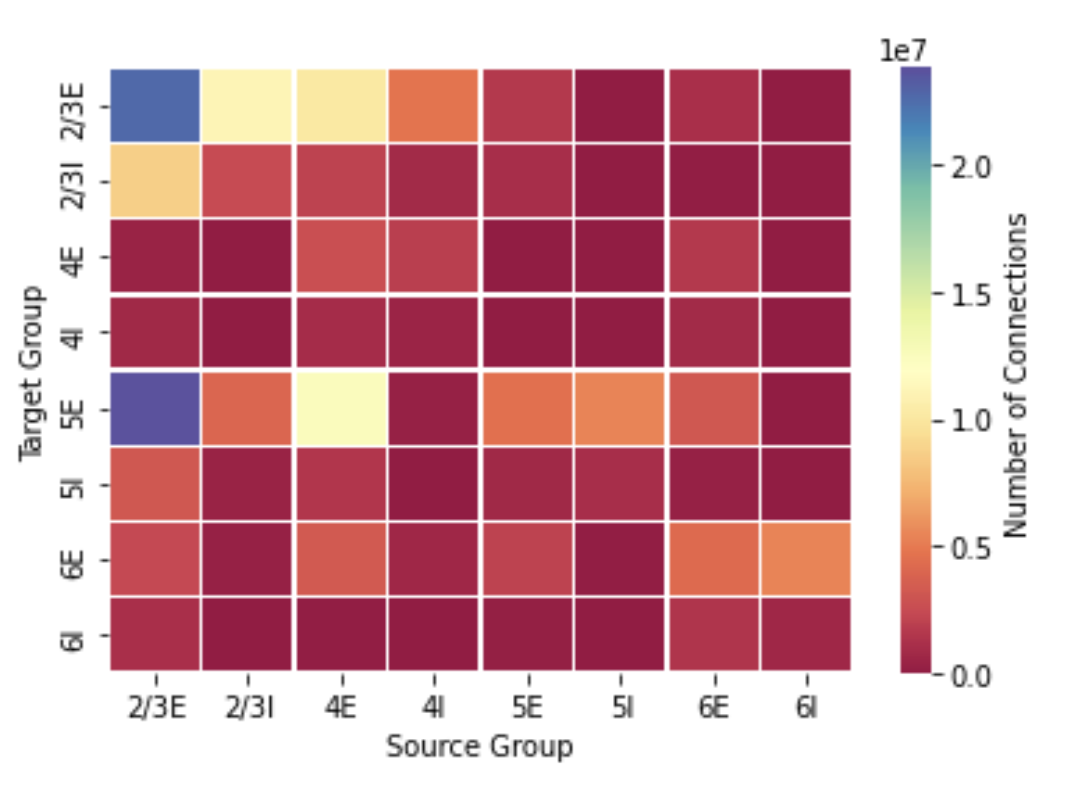}
\caption{Connectivity matrix of neuron populations in the model}
\label{fig:connmatrix}
\end{figure}

Based on Peter's rule and probabilities derived from anatomical and physiological studies, the number of connections ($K$) was calculated for neuron groups using equation~\ref{eq:syn_num} (Equation 3 in \citet{Shimoura2018}) where $C_{a}$ is the connection probability defined in table \ref{table:connectivity}, and $N_{pre}$ and $N_{post}$ are the sizes of the presynaptic and postsynaptic populations, respectively. The $C_{a}$ values in the connectivity matrix (Table \ref{table:connectivity}) describing the probabilities of connections between groups were adapted from the original \citet{Potjans2014} study in this motor cortex model based on changes to the number of neurons in each layer, while maintaining the average relative number of connections in each neuron target group as the original model. The resulting connectivity was similar to experimental investigations of layer specific wiring in the motor cortex \citep{Weiler2008}, notably the dominant layer 2/3 to layer 5 connection pathway (Figure~\ref{fig:connmatrix}). 

\begin{equation}
\label{eq:syn_num}
K = \frac{log\left(1 - C_{a}\right)}{log\left(1 - \frac{1}{\left(N_{pre}N_{post}\right)}\right)}
\end{equation}


\begin{table}[bt]
\centering
\caption{Connectivity probability values ($C_a$) used in equation~(\ref{eq:syn_num}) to determine the number of synapses (K) between each group.}
\label{table:connectivity}
\begin{threeparttable}
\begin{tabular}{ccccccccc}
\headrow
\multicolumn{9}{c}{\textbf{Source Group}}\\
\headrow
\thead{Target Group} & \thead{2/3E} & \thead{2/3I} & \thead{4E} & \thead{4I} & \thead{5E} & \thead{5I} & \thead{6E} & \thead{6I}\\
\textbf{2/3E} & 0.192 & 0.3095 & 0.3356 & 0.5802 & 0.0143 & 0 & 0.0159 & 0 \\
\textbf{2/3I} & 0.252 & 0.2553 & 0.2558 & 0.4183 & 0.034 & 0 & 0.008 & 0 \\
\textbf{4E} & 0.016 & 0.012 & 0.3725 & 0.7704 & 0.0031 & 0 & 0.0879 & 0 \\
\textbf{4I} & 0.1334 & 0.006 & 0.5266 & 0.8295 & 0.0013 & 0 & 0.2007 & 0 \\
\textbf{5E} & 0.1902 & 0.1202 & 0.3785 & 0.0592 & 0.0377 & 0.1662 & 0.0396 & 0 \\
\textbf{5I} & 0.1071 & 0.0533 & 0.2129 & 0.0201 & 0.027 & 0.1374 & 0.0179 & 0 \\
\textbf{6E} & 0.0318 & 0.014 & 0.1754 & 0.1597 & 0.0257 & 0.0078 & 0.0784 & 0.399 \\
\textbf{6I} & 0.0708 & 0.002 & 0.0269 & 0.0101 & 0.0125 & 0.0031 & 0.1276 & 0.267 \\
\hline  
\end{tabular}
\begin{tablenotes}
\item E, excitatory; I, inhibitory. Number indicates cortical layer.
\end{tablenotes}
\end{threeparttable}
\end{table}

\begin{figure}[htbp!]
\centering
(a)\includegraphics[scale=0.3]{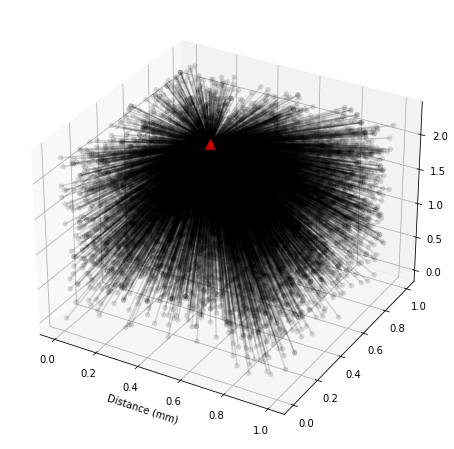}
(b)\includegraphics[scale=0.3]{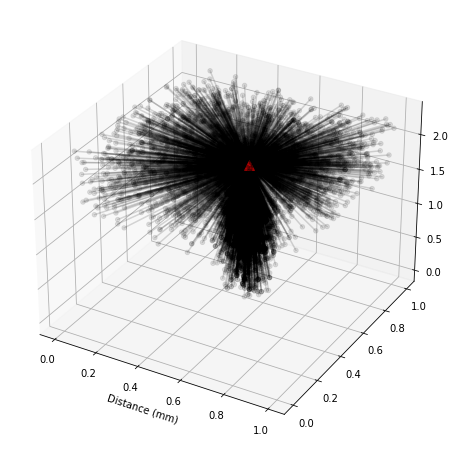}
\caption{Connectivity of a single excitatory neuron on layer 2/3 (a) random connectivity (b) local connectivity. Model represents 1$mm^2$ of cortical surface. Units along all axes are in mm.}
\label{fig:neuronconnectivity}
\end{figure}

The original \citet{Potjans2014} cortical model implemented a random connectivity scheme. This was reimplemented and adapted here to model the motor cortex, as described above, and a new local connectivity scheme was developed. In the random and local connectivity schemes, the total number of synapses in both models was kept constant for fair comparison. Neurons were given spatial parameters (X, Y, and Z coordinates) to be distributed over a $1~\text{mm}^2$ surface and a depth range based on the group's layer. The likelihood of a neuron connecting with another neuron was determined by a Gaussian distribution over distance, where $x$ and $y$ are a neuron's spatial location and the $radius$ parameter defines the extent of the spread (Equation \ref{eq:connection_prob}). This connectivity is compatible with the synapse definition of the original random model described earlier but the added spatial description of neurons allowed constraining the synapses to a localised connectivity. The spatial connectivity description was also adapted from previous cortical modelling work \citep{Lumer1997, Hill2004, Esser2005}. Intralaminar connectivity had a greater radius of connectivity than interlaminar connectivity and inhibitory connections had a lower connectivity radius than excitatory connections as supported by physiological data \citep{Bender2003, Boucsein2011, Fino2013, VanPelt2013}. The radius value for different connection types is given in table \ref{table:Radius}. Multiple connections could be established between two neurons in both the random and local connectivity schemes and self connections in the local connectivity model were not allowed. Figure~\ref{fig:neuronconnectivity} shows the postsynaptic connections of a single neuron (red) on the random connectivity model and the connectivity of a single neuron in the distance based connectivity model.

\begin{equation}
\label{eq:connection_prob}
Probability = exp\left(-\frac{(x_{pre} - x_{post})^2 + (y_{pre} - y_{post})^2}{2*radius^2}\right)
\end{equation}


\begin{table}[bt]
\centering
\caption{Connection radius for various connection types}
\label{table:Radius}
\begin{threeparttable}
\begin{tabular}{cc}
\headrow
\multicolumn{1}{c}{\textbf{Connection Type}} & \thead{Radius ($\mu m$)}\\
Horizontal Intralaminar Excitatory Connections (Layer 2/3, 4 \& 5) & 300 \\
Horizontal Intralaminar Excitatory Connections (Layer 6) & 225 \\
Vertical Interlaminar Excitatory Connections & 50 \\
Inhibitory Connections & 175 \\
\hline  
\end{tabular}
\end{threeparttable}
\end{table}

External input was applied to each neuron group as Poisson spike trains at a frequency of 8~Hz. The number of these inputs for each neuron was constant at 2000 inputs for excitatory neuron groups and 1850 inputs for inhibitory neuron groups. This followed the layer-independent input protocol from the original \citet{Potjans2014} model. A sensitivity analysis on the effect of the input on the resulting firing rates was also performed by manipulating the number of inputs and frequency.

To characterise the activity of the network, spike trains and firing rates of neurons and neuron groups were monitored. A spike was defined by the neuron reaching the threshold value of -50~mV. Firing rates were calculated by counting the total number of spikes in each time step of the simulation (0.1~ms). Simulations were run for 500~ms and the first 50~ms was ignored to allow the model to reach steady state. The interspike intervals (ISIs) for each neuron were calculated as the time between spikes. Irregularity was characterised by the coefficient of variation (CV) of the ISI distribution. CVs were calculated as the ratio of the standard deviation to the mean ISIs (Equation \ref{eq:cv}) for each individual neuron in the population, as defined previously \citep{Shimoura2018, Potjans2014}. Firing rates and CVs reported are the mean across neurons in the neuron group, and over time. Notably, firing rates and CVs closely matched values reported by the original model when the settling period was included, consistent with the analyses conducted by \citet{Shimoura2018}, but due to the inconsistency in the spiking behaviour in the first $50$~ms, this settling period was excluded for the analyses carried out in this study.

\begin{equation}
\label{eq:cv}
CV = \frac{\sigma_{ISI}}{\mu_{ISI}}
\end{equation}
\section{Results} \label{results}

\begin{figure}[htbp!]
\centering
(a)\includegraphics[scale=0.27]{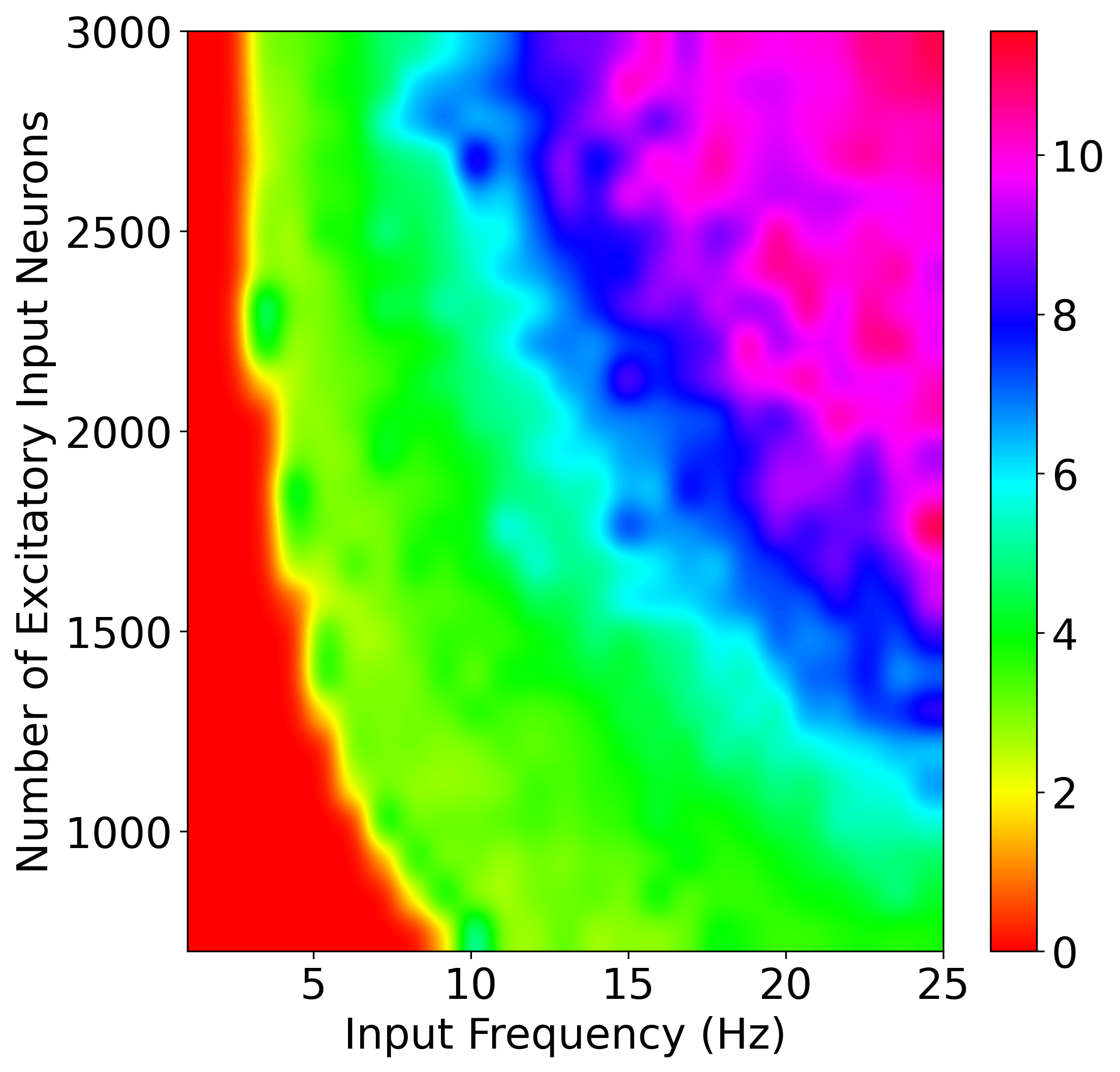}
(b)\includegraphics[scale=0.27]{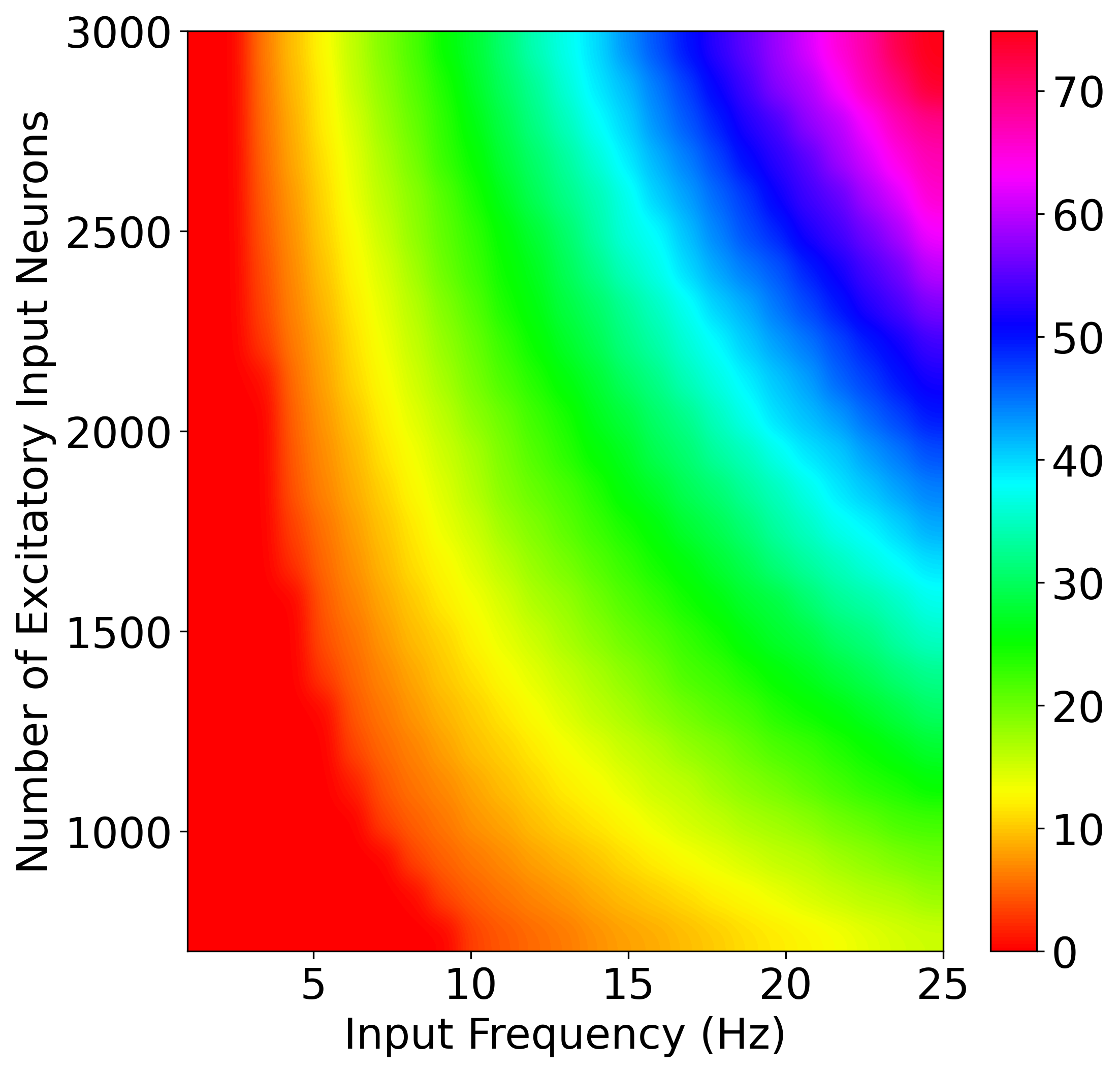}
\caption{Average firing rates in model with changes to input number and frequency for (a) random connectivity and (b) local connectivity. Inhibitory neuron inputs were kept proportional at 92.5\% of excitatory inputs. Scales represent average firing rate across all neurons. Note the different range in scales of the colour maps.}
\label{fig:input}
\end{figure}

The motor cortex receives connections from multiple brain regions such as the thalamus, premotor cortex and somatosensory cortex; this complexity makes selection of an appropriate surrogate input challenging. The effect of the input number and frequency to the model was explored while keeping the ratio of excitatory and inhibitory input and relative input between the different neuron groups constant (ie. 2000 for excitatory neurons, 1850 for inhibitory neurons). A single input frequency was used across the model though physiologically there may be variation in the number of inputs and range of input frequencies. Networks with local connectivity show greater sensitivity to the frequency and number of inputs, with a greater range of resultant average firing rates. Figure~\ref{fig:input} shows a heat map of the effect of input number and frequency on the average firing rates of neurons in the model, note the change in scale in each colour map. 

\begin{figure}[htbp!] 
\centering
(a)\includegraphics[scale=0.3]{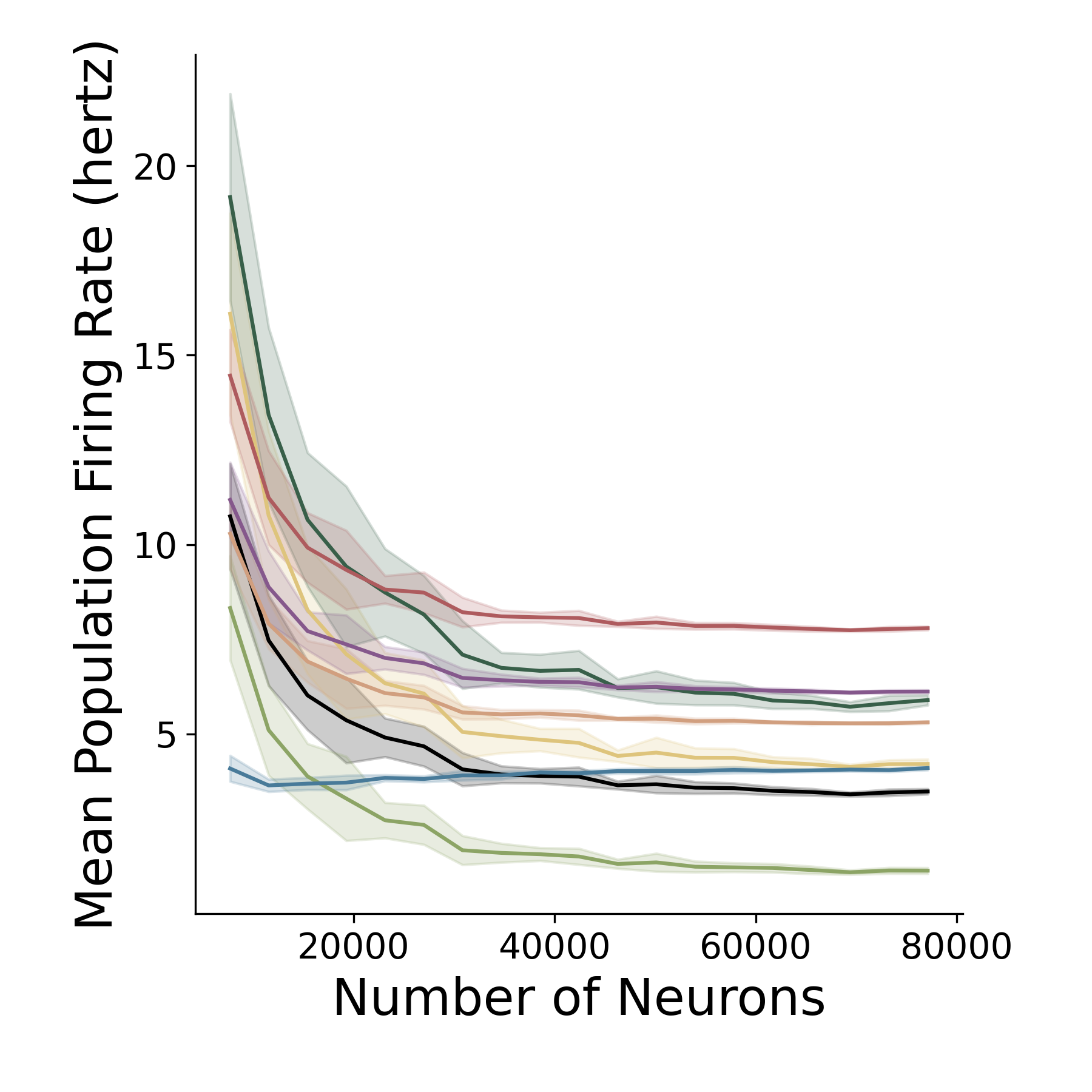}
(b)\includegraphics[scale=0.3]{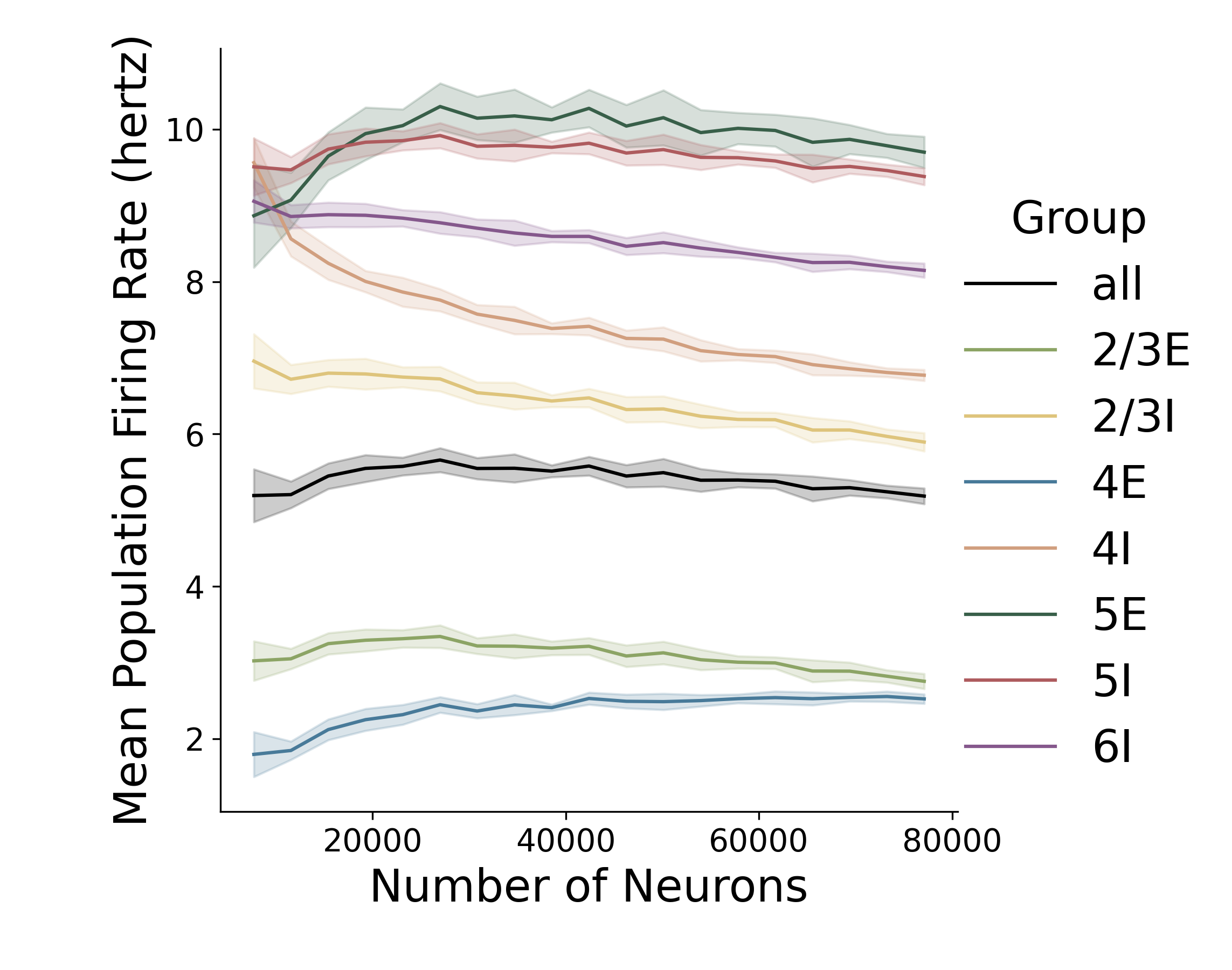}
(c)\includegraphics[scale=0.3]{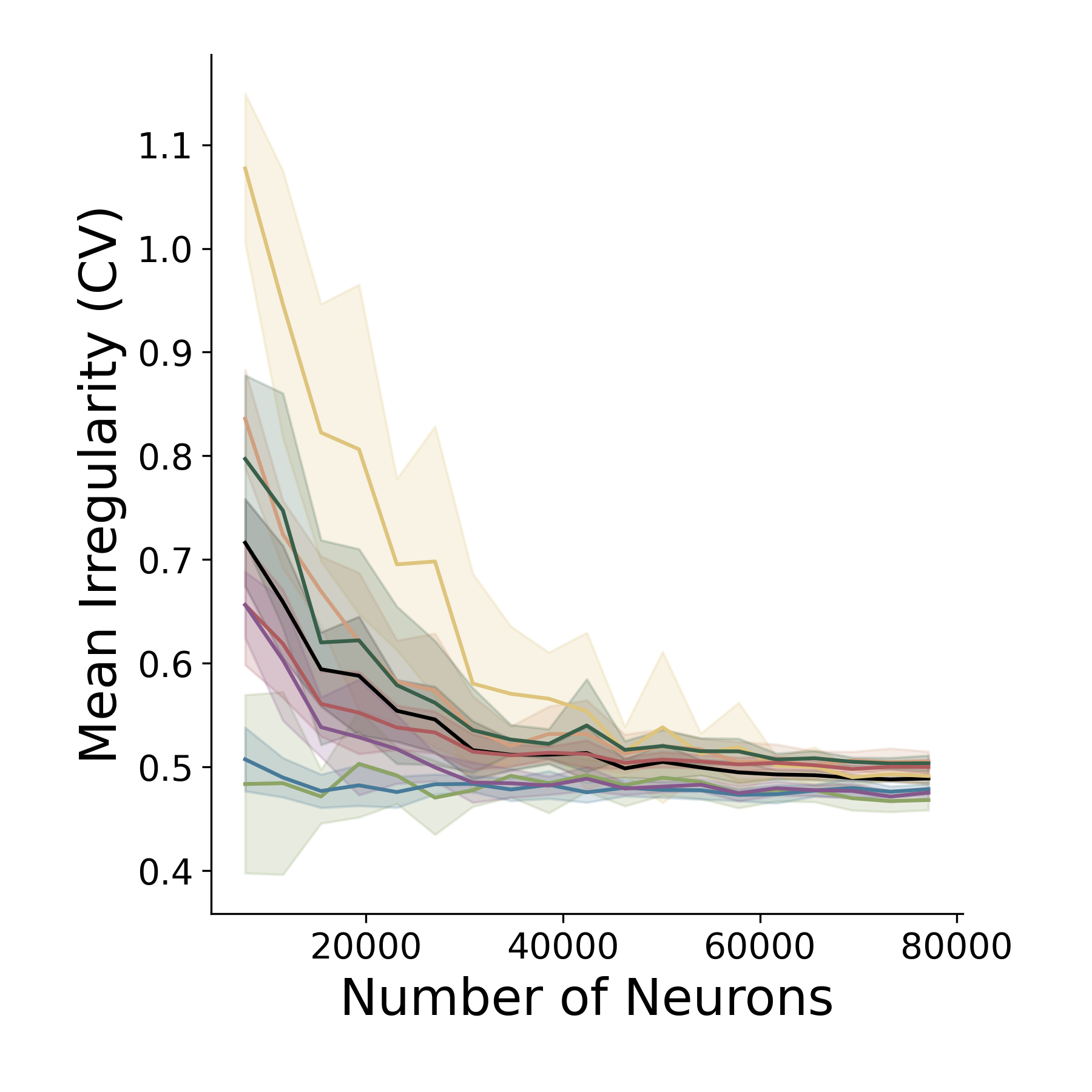}
(d)\includegraphics[scale=0.3]{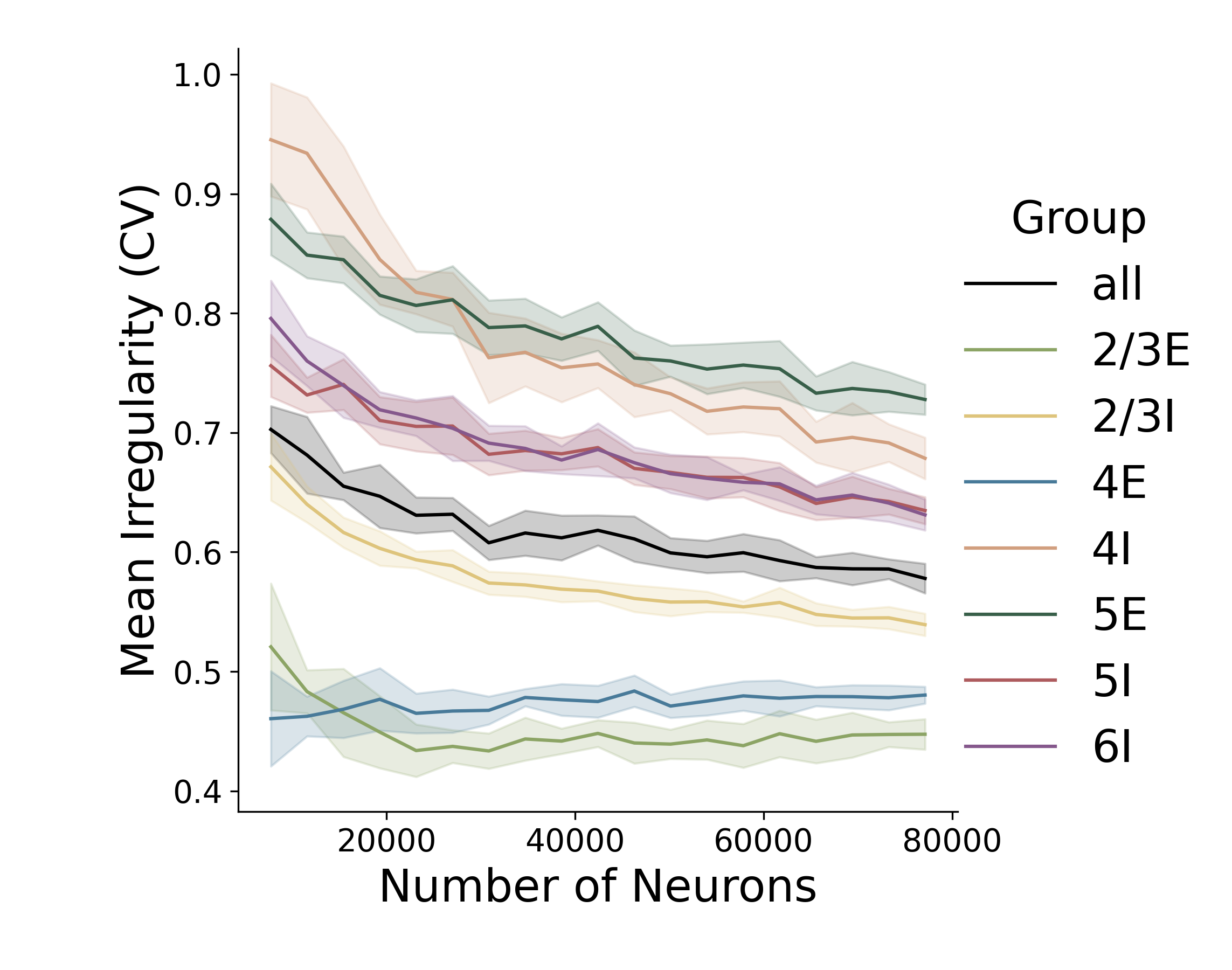}
\caption{Numerical sensitivity analysis for firing rates of neuron groups in (a) random connectivity and (b) local connectivity, and the coefficient of variation in neuron groups with (c) random connectivity model of the motor cortex and (d) local connectivity. The proportions of each neuron group and relative number of synapses were kept the same.}
\label{fig:numsensitivity}
\end{figure}

An analysis of the effect of the number of neurons in the model showed that consistent frequencies and CVs were produced by a random network containing 35,000 or more neurons. Neuron counts by isotropic fractionator and flow fractionator methods of the motor cortex in old and new world primate species lie in the estimated range of 50,000--90,000 neurons per $\text{mm}^2$ \citep{Young2013, Collins2016}. Figure~\ref{fig:numsensitivity} shows the firing rates and CVs in the network over a range of 0--80,000 neurons for both random and local connectivity, with standard deviation measures over ten simulations. The firing rates and CVs were more stable in the local connectivity model at lower number of neurons. In both models, layer 6E neuron CVs were not able to be calculated or were highly variable, probably due to very low firing rates in the neuron group, and were thus excluded from the figures. 

The firing rates of the model were within an appropriate range of baseline physiological recordings of a number of studies, with lower firing rates in layer 2/3 and layer 6 of the model than layer 5 \citep{Crochet2009}. \citet{Beloozerova2003} presented microelectrode measurements of the motor cortex in awake rabbits, with mean discharge rates of 0.6~Hz in corticocortical neurons, prevalent in layer 2/3, 5.7~Hz in layer 5 neurons and 0.4~Hz in layer 6 neurons. \citet{Schiemann2015} published similar values from whole cell patch clamp recordings in mice. \citet{Dabrowska2021} recorded firing rates in two awake monkeys and firing rates in a resting state were 6.3~Hz and 7.81~Hz and CV measures of 0.83 and 0.79 which lies closer to the irregularity of a random poisson distribution (CV = 1). These were recorded using a 100 electrode Utah Array in the hand motor area of macaque monkeys, at a depth of 1.5~mm, which would correspond to deep layer 3 or superficial layer 5 \citep{Ninomiya2019}. \citet{Lacey2014} reported a similar mean value of 7.1~Hz firing in layer 5 motor cortex cells at rest, from micro-electrode recordings in rats. CVs of spiking activity in cortical neurons have been consistently reported in previous literature as being close to 1 \citep{Noda1970, Holt1996}. 


\begin{table}[bt]
\centering
\caption{Mean firing rates and coefficient of variability, with standard deviations, of neuron group activity in random and local connectivity schemes.}
\label{table:ModelFiringRates}
\begin{threeparttable}
\begin{tabular}{ccccccccc}
\headrow
{} &  \multicolumn{4}{c|}{\textbf{Random Connectivity}} & \multicolumn{4}{c|}{\textbf{Local Connectivity}}\\
\headrow
\begin{tabular}{@{}c@{}}\textbf{Neuron} \\ \textbf{Group}\end{tabular} & \begin{tabular}{@{}c@{}}\textbf{Mean FR} \\ \textbf{(Hz)}\end{tabular} & \begin{tabular}{@{}c@{}}\textbf{STD} \\ \textbf{FR}\end{tabular}  & \begin{tabular}{@{}c@{}}\textbf{Mean} \\ \textbf{CV}\end{tabular} & \begin{tabular}{@{}c@{}}\textbf{STD} \\ \textbf{CV}\end{tabular}  & \begin{tabular}{@{}c@{}}\textbf{Mean FR} \\ \textbf{(Hz)}\end{tabular} & \begin{tabular}{@{}c@{}}\textbf{STD} \\ \textbf{FR}\end{tabular}  & \begin{tabular}{@{}c@{}}\textbf{Mean} \\ \textbf{CV}\end{tabular} & \begin{tabular}{@{}c@{}}\textbf{STD} \\ \textbf{CV}\end{tabular}\\
\hline
2/3E & 1.86 &  8.44 & 0.51 & 0.27 & 3.24 & 6.98 & 0.42 & 0.21 \\
2/3I & 4.81 & 11.78 & 0.56 & 0.26 & 6.57 & 8.81 & 0.57 & 0.25 \\
4E & 3.99 & 7.44 & 0.48 & 0.24 & 2.55 & 5.25 & 0.48 & 0.23 \\
4I & 5.51 & 12.97 & 0.51 & 0.24 & 7.42 & 14.71 & 0.75 & 0.28 \\
5E & 6.90 & 15.16 & 0.58 & 0.24 & 10.51 & 10.87 & 0.79 & 0.27 \\
5I & 8.13 & 9.22 & 0.51 & 0.22 & 9.90 & 10.49 & 0.67 & 0.26 \\
6E & 0.008 & 0.034 & - & - & 0.125 & 0.59 & 0.55 & 0.27 \\
6I & 6.42 & 9.10 & 0.51 & 0.22 & 8.66 & 10.96 & 0.67 & 0.25 \\
\hline  
\end{tabular}
\begin{tablenotes}
\item FR, Firing Rates; STD, Standard Deviation; CV, Coefficient of Variability; E, Excitatory; I, Inhibitory
\end{tablenotes}
\end{threeparttable}
\end{table}

Based on these experimentally reported values, the firing rates of the neural network model with random connectivity were closer to the physiological range, though with slightly higher firing in layer 2/3 and lower in 6. The local connectivity model showed an increase in mean firing rates over all the neuron populations and a wider range in mean CV. The mean CV values in the local connectivity model more closely matched experimental recordings in the middle-deep layers, notably in 5E which is the dominant neuron group. Firing frequencies were also observed to be higher in inhibitory neuron groups. Table~\ref{table:ModelFiringRates} reports the mean values and standard distributions of firing rates and irregularity (CV) in the random connectivity model and local connectivity model.

\begin{figure}[htbp!]
\centering
(a)\includegraphics[scale=0.27]{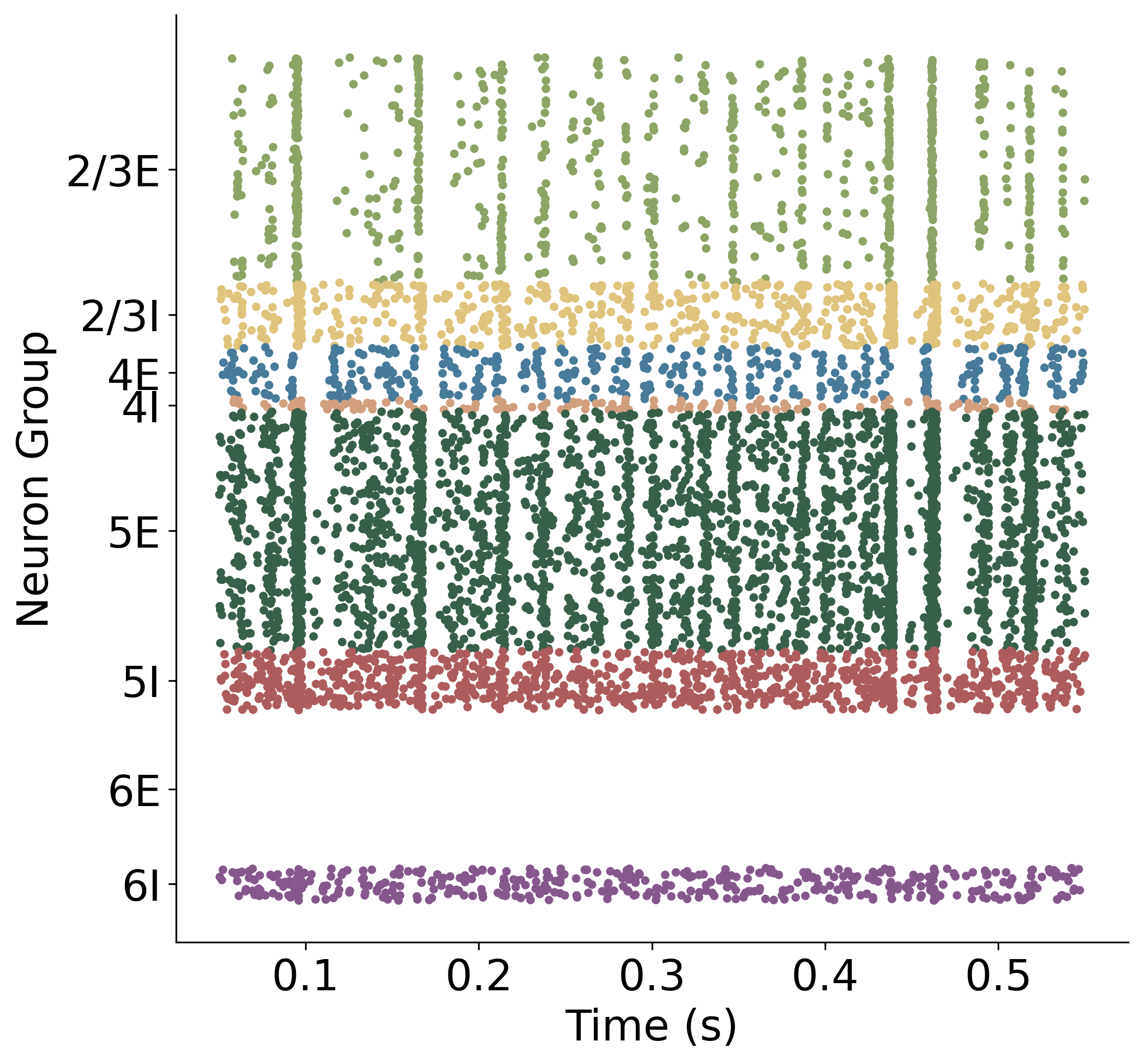}
(b)\includegraphics[scale=0.27]{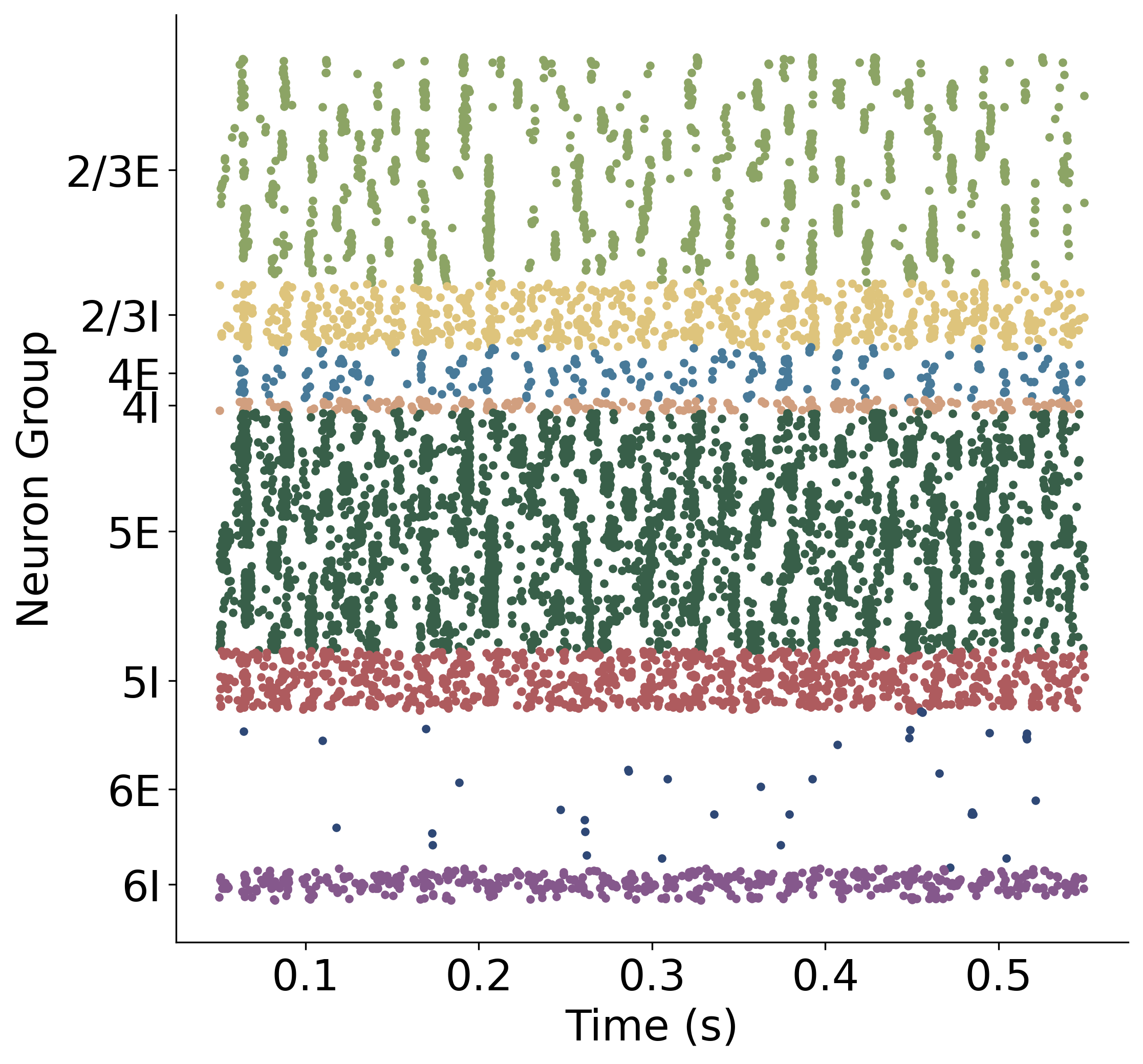}
(c)\includegraphics[scale=0.27]{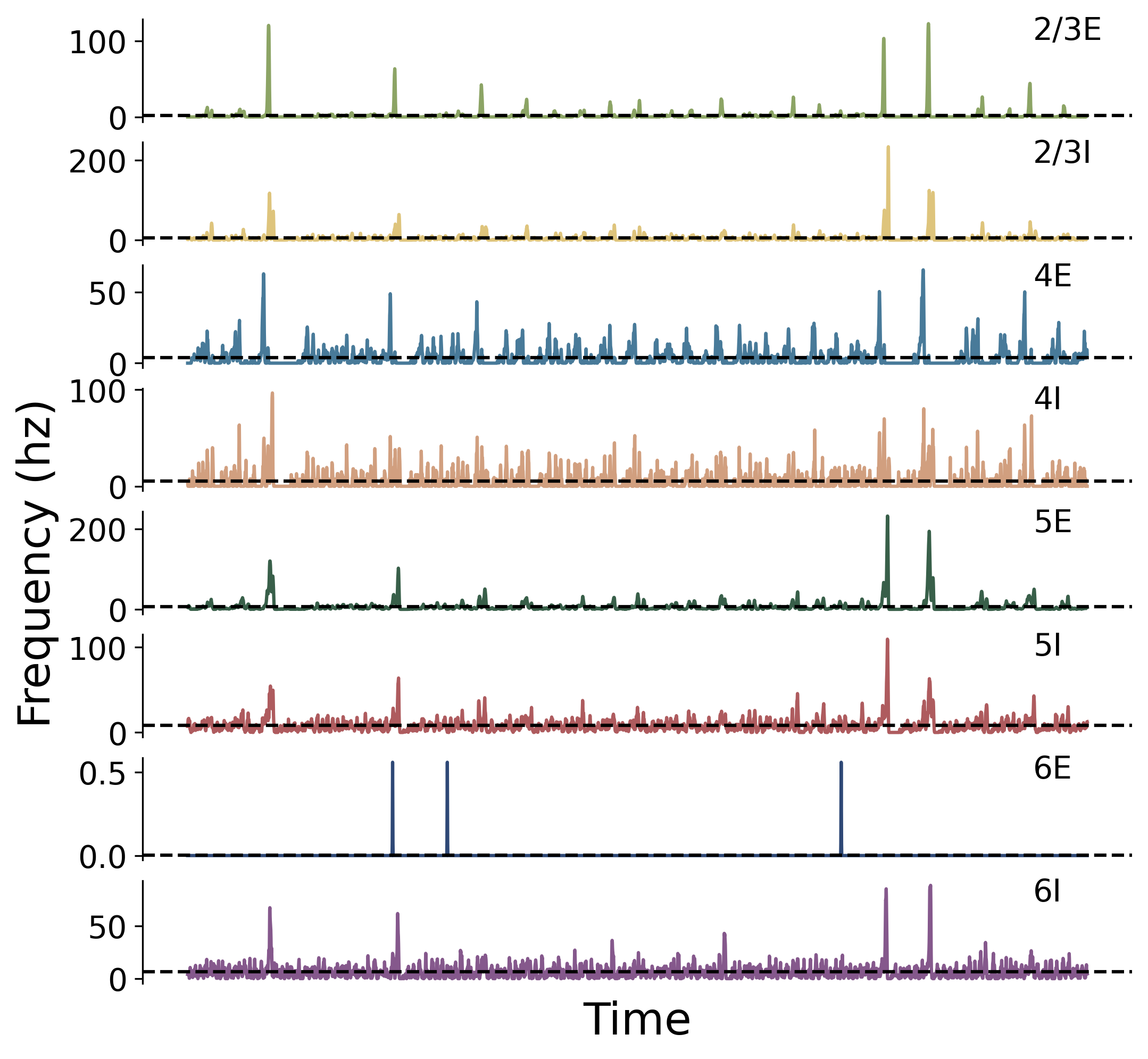}
(d)\includegraphics[scale=0.27]{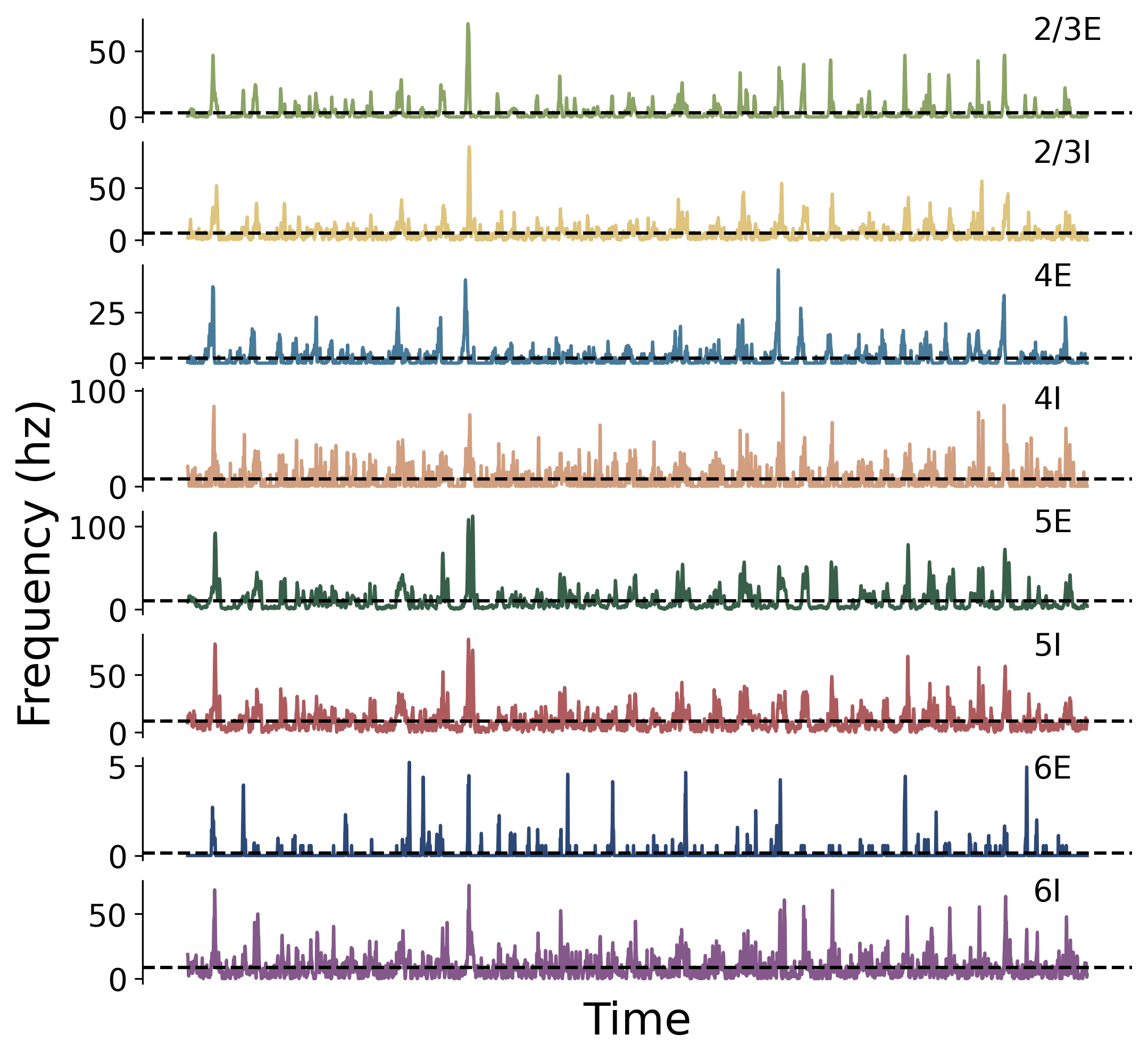}
\caption{Raster plots of the model spiking behaviour with (a) random connectivity and (b) local connectivity. Population rates of model with (c) random connectivity and (d) local connectivity.}
\label{fig:rasterfiring}
\end{figure}

The activity of neurons in the model showed a different pattern of activity in the local connectivity model compared to the random connectivity. Figure~\ref{fig:rasterfiring} shows a raster plot of 10\% of the neurons in the model and a plot of the frequencies across the neuron groups. Firing rates of the neuron groups generally remained $<\ $100~Hz but the random connectivity model showed a greater tendency for sudden high frequency firing activity across the network. This synchrony of firing activity could be neuronal avalanches, bursts of firing, which are a noted feature of brain dynamics and indicator of scale-free critical dynamics in complex systems \citep{Arviv2019}. 

\begin{figure}[htbp!]
\centering
(a)\includegraphics[scale=0.35]{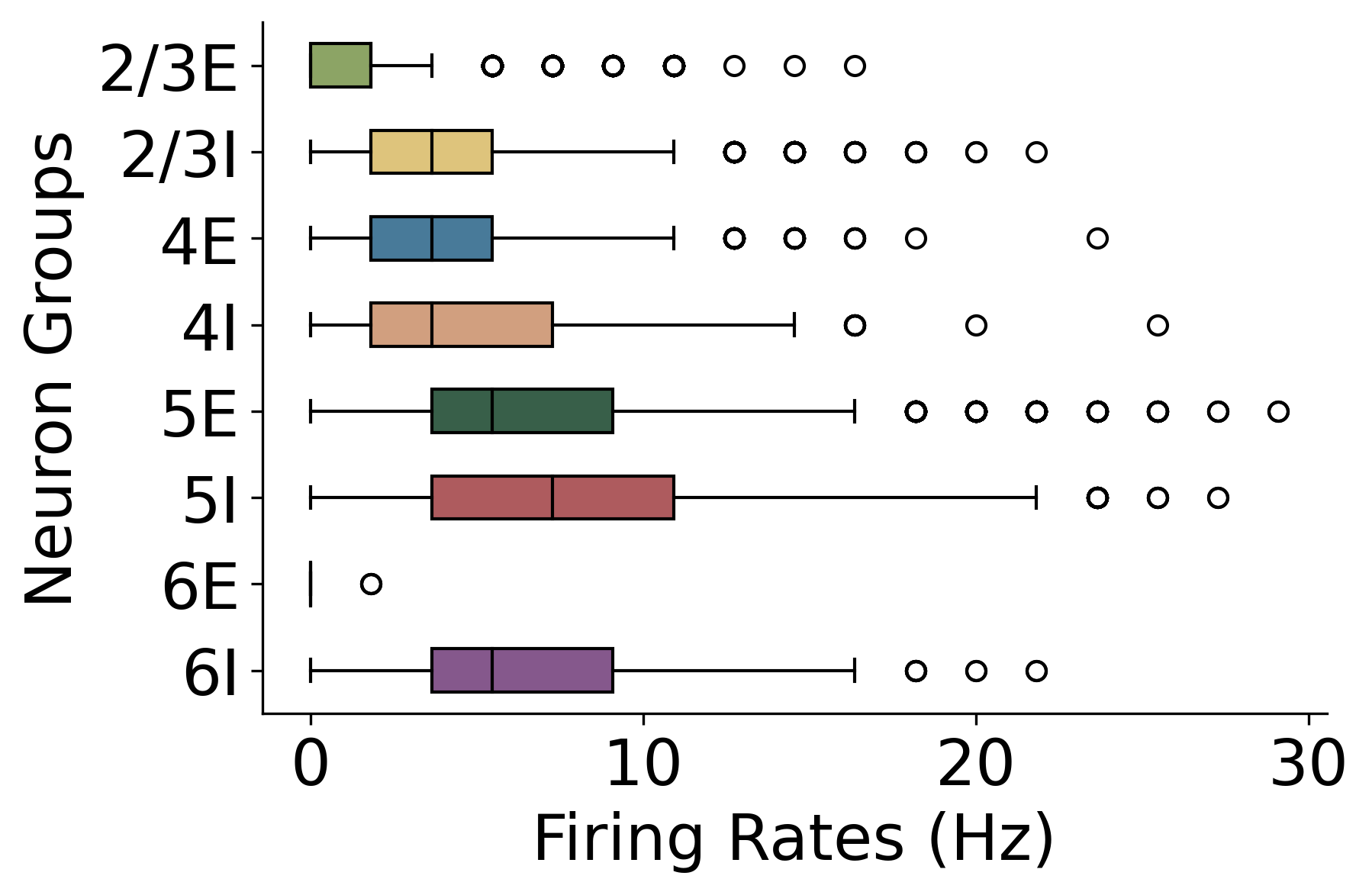}
(b)\includegraphics[scale=0.35]{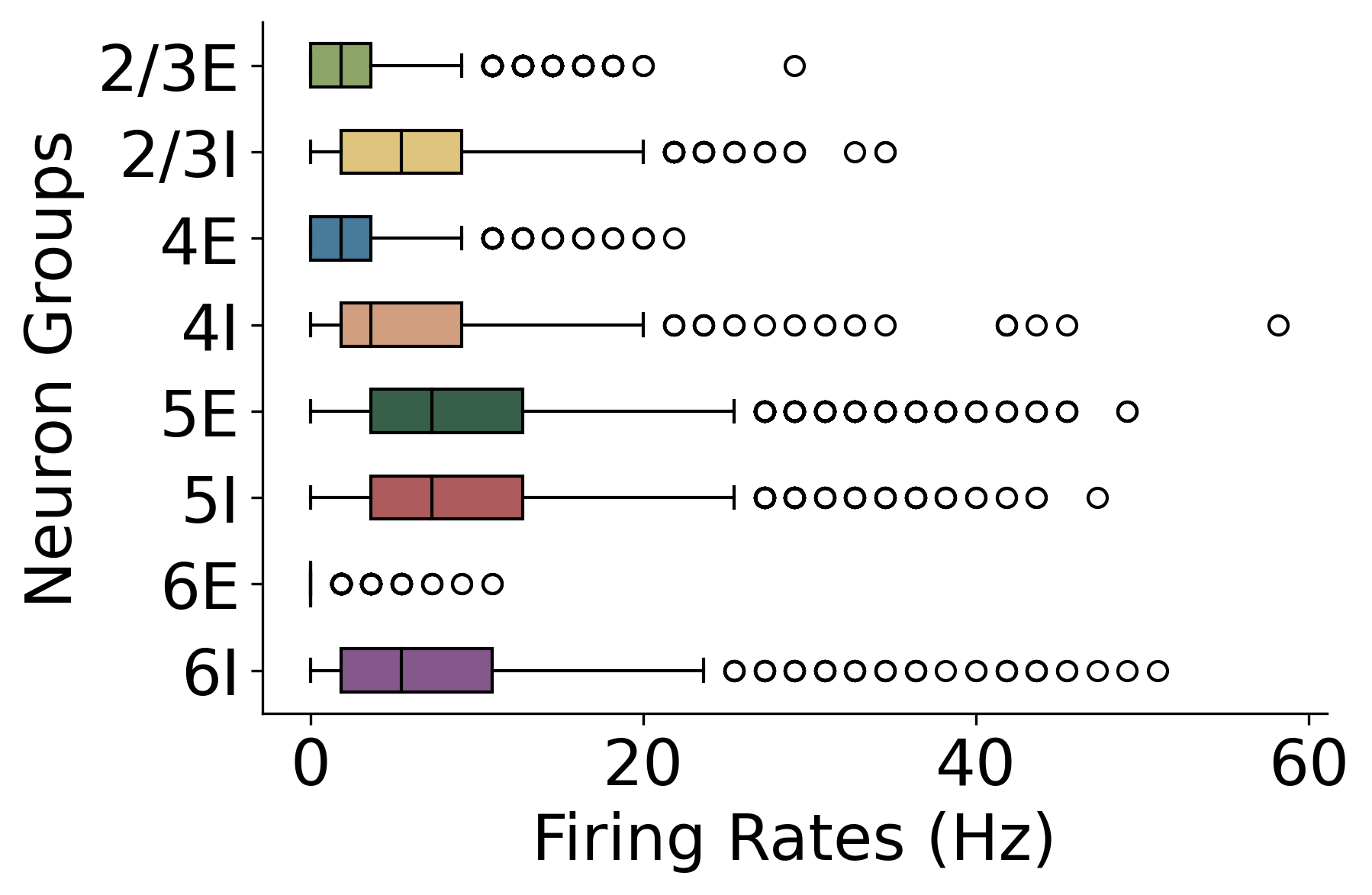}
(c)\includegraphics[scale=0.35]{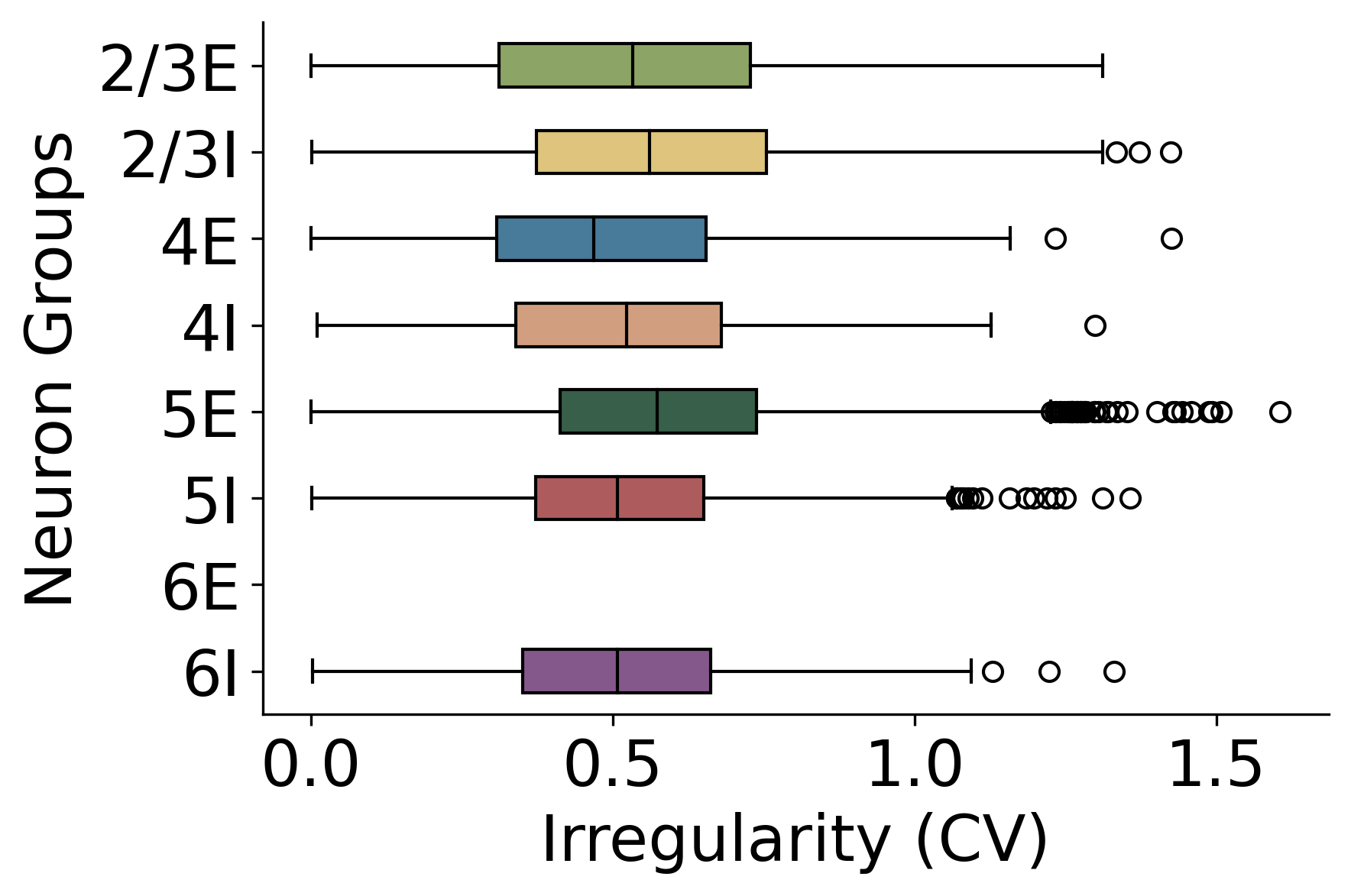}
(d)\includegraphics[scale=0.35]{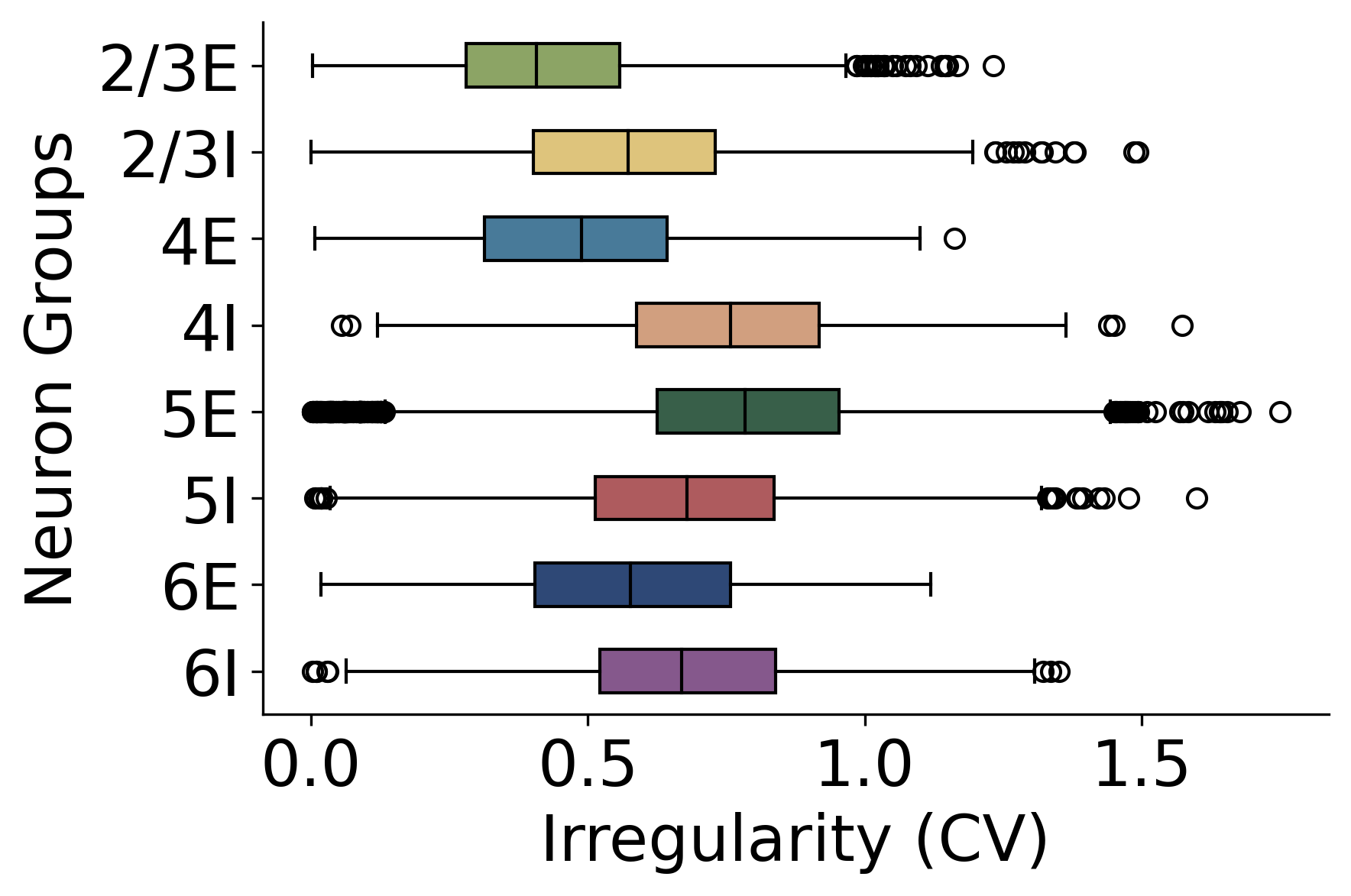}
\caption{Firing rates box and whiskers for (a) random connectivity and (b) local connectivity. Irregularity (CV) box and whiskers plot for (c) random connectivity and (d) local connectivity.}
\label{fig:boxplots}
\end{figure}

The range of firing rates in recorded neurons also varies with long-tailed distributions of firing rates \citep{Roxin2011}. The range of firing is reported to be 0--40~Hz in measurements of awake mice and cats \citep{Armstrong1984, Estebanez2017}, while recordings in monkeys ranged from 0--100~Hz \citep{Dabrowska2021}. The firing rates of the model also showed long tailed distributions (Figure~\ref{fig:boxplots}), with longer tailed distributions, closer to physiological data, in the local connectivity model. 

\begin{figure}[htbp!]
\centering
\includegraphics[scale=0.4]{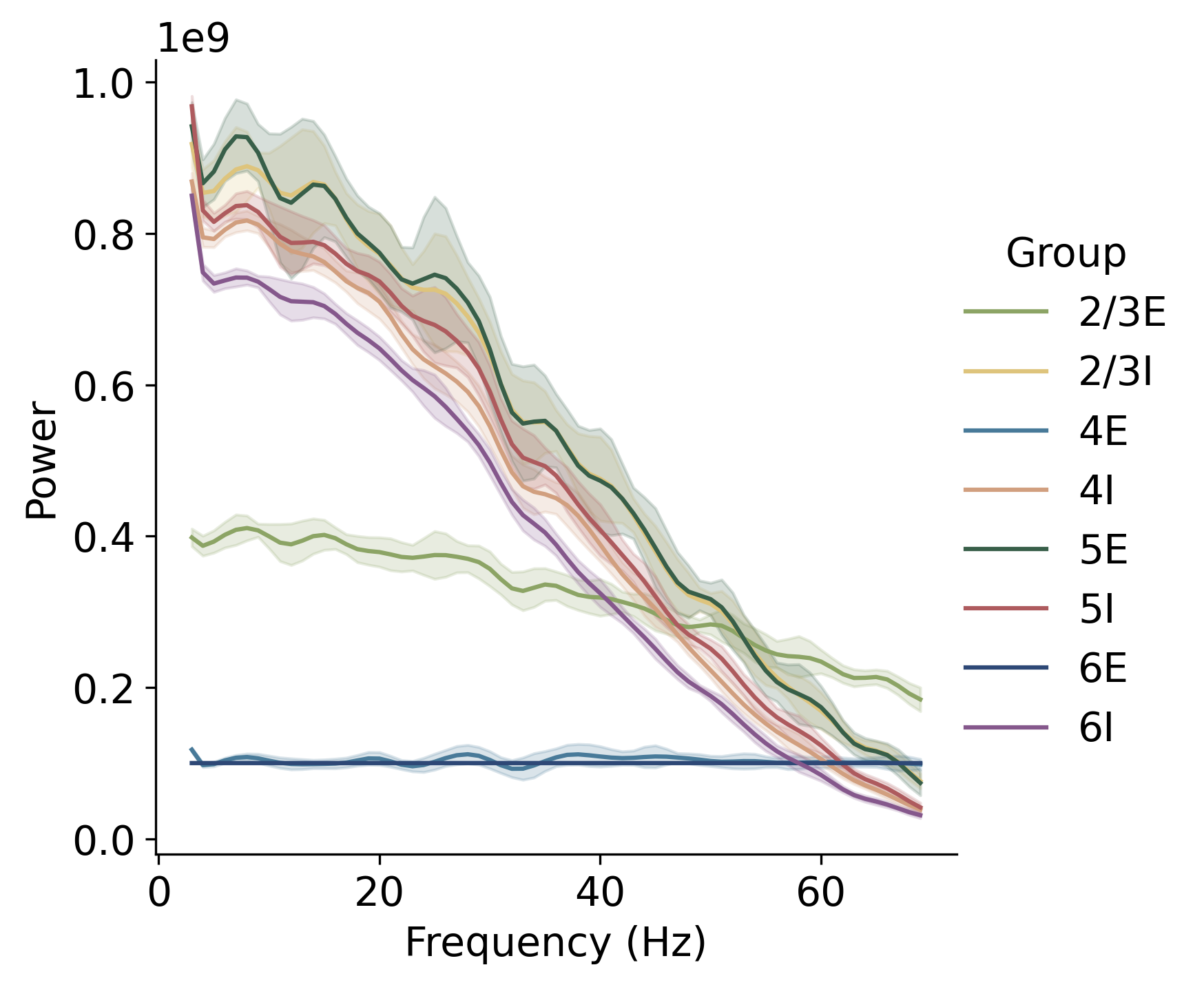}
\includegraphics[scale=0.4]{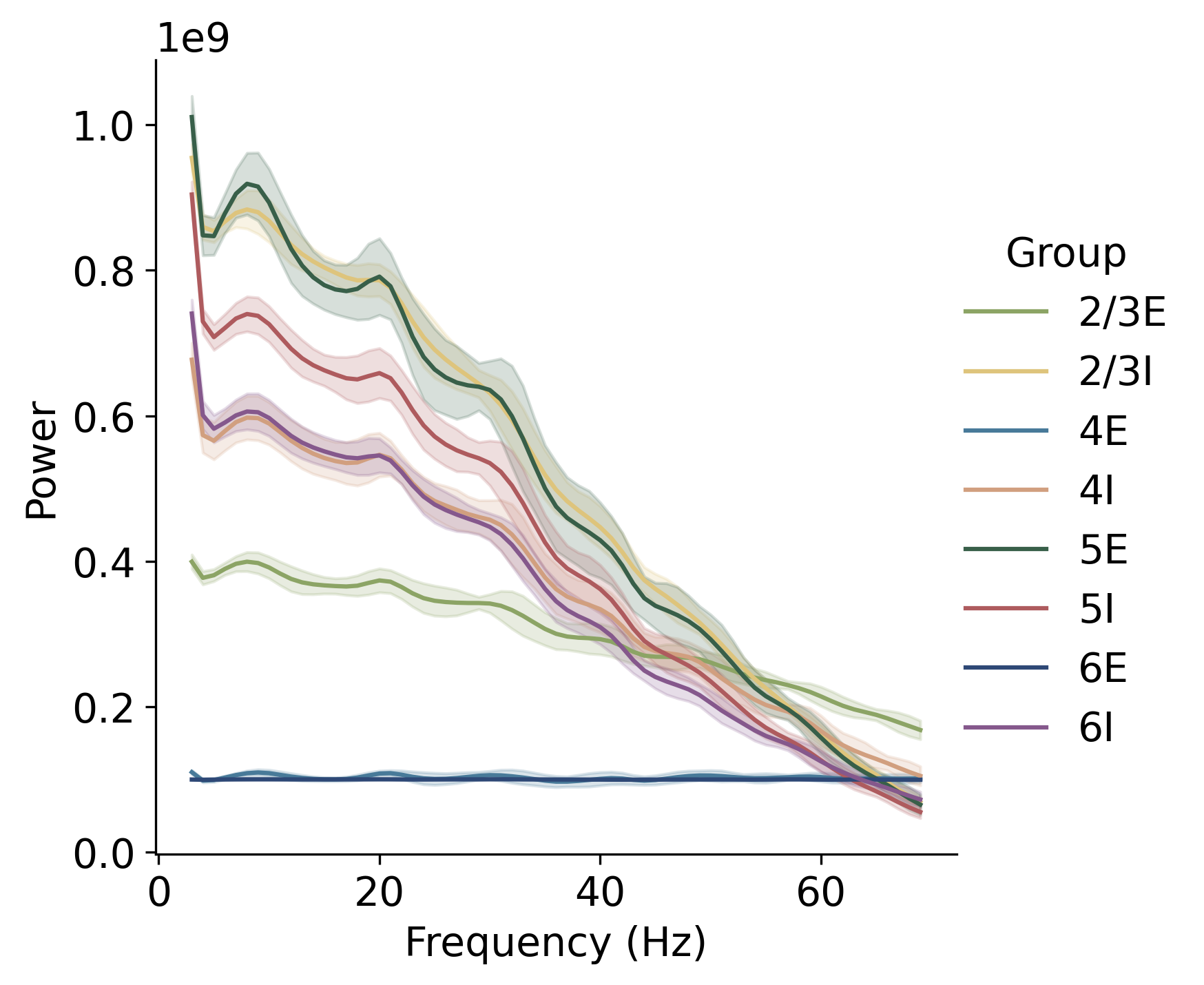}
\caption{Power spectrum of random (left) and local (right) connectivity with observed beta 'peaks' between $10 - 30$Hz.}
\label{fig:power}
\end{figure}

A power spectrum analysis shows a peak between 10--30~Hz which lies within the beta oscillation range (13--30~Hz). Experimentally, beta waves are observed by EEG, magnetoencephalography (MEG) or electrocorticography (ECoG), though the origin and role of this rhythm is still unclear \citep{Jensen2005, Khanna2017, Kilavik2013, Barone2021}. This model suggests that oscillatory rhythms could be emergent properties of the cortical network since the input to this model is non-oscillatory. The standard deviation, taken over ten simulations, was slightly reduced in the local connectivity model suggesting a potential regularisation of the power spectrum and beta peak (Figure~\ref{fig:power}). 

\section{Discussion} \label{discussion}
This study builds on previous cortical modelling work, and investigates the effects of a physiologically-realistic, spatially defined local connectivity scheme on neuron activity, while focusing on replicating the spontaneous firing in the motor cortex. We replicated spontaneous firing behaviour in regards to neuron firing rates, irregularity, and power spectrum peaks, as recorded by previous experimental data. Local and random connectivity was compared in the model to provide insight to the topological network properties that influence population-based firing behaviour. Local connectivity showed qualitatively different temporal patterns of firing and more realistic irregularity (CV). It also resulted in a wider tailed distribution of firing and a narrower standard deviation in the power spectrum. 

Spontaneous activity in the cortex has been characterised by low firing rates and asynchronous activity. The model with random connectivity replicated previously measured values with lower firing rates in layers 2/3 and 6 and higher firing rates in layer 5. The increase in firing rates in the local connectivity model may be due to more coordinated inputs within a local cluster resulting in a greater likelihood to reach threshold. This could be broadly explained by the theory of neurons as coincidence detectors, which states that information is encoded and propagated by the timing of action potentials \citep{Softky1993, Shadlen1994, Chen2013, deCharms1998}. Though local connectivity increased the firing rates of the network, similar proportions of firing rate values between layers were maintained, which suggests additional tuning of input or overall connectivity density might resolve the differences in absolute firing rate and be able to reduce firing to match experimental results. 

The local connectivity model replicated experimentally derived CV values in the deeper layers of 5 and 6, with similar CV values to the random connectivity model in the superficial layers 2/3 and 4. As such, the topological structure played a key role in the variability of firing activity. The local connectivity model also exhibited a lower standard deviation in the power spectrum, compared to the random connectivity model. The narrower range in the power spectrum suggests a more reliable occurrence of beta wave frequencies that are consistently observed in EEG recordings. The more realistic, local connectivity model could therefore be a topological pattern in the cortex which contributes to the occurrence of the beta wave. 

Compared to the random connectivity model, the local connectivity scheme showed more sensitivity and instability in response to input activity. Within the range of changes in the input, the local connectivity model covered a larger range of frequencies in the firing behaviour of the network. The greater sensitivity to input could be a key factor in the highly variable activity in the cortex which might be necessary for phase transitions or changes in state, a notion described by criticality \citep{Beggs2012, Arviv2019, Chialvo2010, Deco2012}. This critical state is hypothesised to be functionally beneficial for efficient information transmission in the cortex \citep{Shew2013}. In the motor cortex, being in the critical state may play a role in the generation of a wide range of voluntary muscle movements. 

The E/I ratio is critical in determining the model dynamics and emergent behaviour \citep{Zhou2018, Deco2014}. The balance of E/I activity is linked to scale-free dynamics and operating near a critical point of activity, which does not extinguish or explode into seizure \citep{Kumar2008, Shew2013}. The overall level of excitation and inhibition, as well as the detailed dynamics of excitatory and inhibitory synaptic conductance, has a large effect on circuit activity. Our model contained 24\% excitatory neurons, with post-synaptic conductance of inhibitory connections modelled with a conductivity 4 times stronger than excitatory connections, resulting in a stable E/I balance. However, this E/I ratio could be an area of further exploration as recently a study by \citet{Bakken2021} looked at rat, marmoset and human cell types and found a proportion of GABAergic neurons as 16\% in mouse primary motor cortex, 23\% in marmosets and 33\% in human primary motor cortex.

The random connectivity model has been used in previous studies of cortical dynamics \citep{VanVreeswijk1996, Brunel2000}. Random networks are less clustered in their connections, have longer range connections, and are considered to be inefficient with regards to information transfer \citep{Westlake2011}. \citet{Mehring2003} implemented a locally connected random network (LCRN), with distance-based connectivity in a large scale neural network model containing 100,000 neurons. \citet{Kumar2008} also used a LCRN to investigate the propagation of synchronous and asynchronous activity. Typically, these models have two populations, excitatory and inhibitory neuron groups with recurrent intra- and inter-group connectivity. Previous investigations of connectivity have been limited to 2D networks which have not aimed to replicate physiological behaviour \citep{Yger2011, Voges2012}. However, to further investigate the spatio-temporal activity of the cortex, particularly with longer range connections, across a cortical network a larger spatial surface area is still needed and this will require increased computing power. 

Whether the connection of neurons in the cortex are specific or random has not yet been fully resolved, though recent studies have suggested that neuron morphology and patterns of recorded activity support a notion of specific connections \citep{Callaway2002, Udvary2022}. Axons do not simply connect to neurons based on spatially overlapped locations but selectively target specific neuron types or groups, for example in layer-specific connectivity patterns \citep{Callaway2002, White2002}. There also appears to be both local and long-range connections in the cortex which is suggested to be efficient in regards to wiring cost and information transfer \citep{Voges2010, Voges2012}. A combination of short and long-range connections were included within the local connectivity model with horizontal, intralaminar connections spanning the surface area of the model, while vertical, interlaminar connections were narrower in radius. 

Over 70 years ago, \citet{Mountcastle1957} made experimental recordings of neurons in the cat somatosensory cortex and proposed an organisational unit of a `vertical group of cells extending through all cortical layers' known as a \textit{cortical column}. Columns are groups of neurons, in the range of 0.5~mm in diameter, tuned to a specific stimulation or attribute in the range of 0.5~mm in diameter, with the arrangement of columns patterned like a 'mosaic' \citep{Horton2005}. However, the function of this columnar organisation remains unclear and there is still no agreed upon function or definition \citep{Haueis2016, Schwalger2017, Horton2005}. In the motor cortex, recordings of neurons show directional tuning during upper limb movements with activity \citep{Georgopoulos1986}. Multi-unit recordings also support functional clusters which exhibit directional tuning, with widths of the tuning field reported in the range of 50-250~$\mu$m in diameter \cite{Amirikian2003, Georgopoulos2007, Hatsopoulos2010}. Individual neurons can make connections outside of columns, with horizontal connections spanning up to 1.5~mm \citep{Boucsein2011}. Although this model only represented 1~mm$^\mathsf{2}$ of cortical surface area, future development could explore an area representation with and without an added constraint to model cortical columns. The topographic structure of the motor cortex and its relation to muscle movement parameters is still undefined and with this model, we can begin to explore these intricate connections in the motor cortex.

The effect of neuron types on cortical dynamics might also be an interesting area for future investigation using this model. Spontaneous activity may physiologically reflect a wider range of neuron types or synaptic dynamics, such as bursting or faster and slower transients, than the simplified excitatory and inhibitory neurons captured in this model \citep{Tan2015, Zeldenrust2018, Masquelier2013, Zhao2020, Tomov2014}. The diversity of inhibitory interneurons also plays a significant role in the modulation of cortical dynamics, though there still currently lacks a clear consensus on the classification of neurons \citep{Markram2004, Isaacson2011}. Though our model incorporated a simplified model of neuron dynamics and input, we showed that more realistic patterns of spontaneous activity can be replicated in a cortical circuit with local connectivity. Incorporation of different cell types would certainly have dynamical consequences at the network level.


Our approach increases the biological plausibility of previous cortical modelling work and contrasts with previous models of the motor cortex which have used continuous-value recurrent neural networks \citep{Sussillo2016, Michaels2019}. We recognise that we have not considered other properties such as neuron types, dendritic processing and synaptic plasticity that are likely to play a role in firing dynamics. Biological learning paradigms such as reinforcement learning with spike-timing dependent plasticity (STDP) have previously been incorporated in spiking neuronal models of motor control, though not specifically replicating motor cortex activity \citep{Spuler2015}. Concurrently, experimental data will also be vital for continued model development and validation, a synergy of computational and experimental techniques will be required to elucidate the complex connectivity of cortical circuits and how it contributes to the generation of dynamic activity.

This work is the beginning of a larger scale exploration of neural control in the motor system with scope to extend the model. The proposed model could be incorporated into feedforward and feedback circuits in the neuromusculoskeletal system involving the spinal cord and alpha-motoneuron pools. Motor-unit and muscle recruitment might be task-specific \citep{terHaar1984}, however, the role of the motor cortex in executing movement commands has not yet fully been elucidated. Our model could be used to explore the generation of muscle activity \citep{Marshall2022}. The model could also fit into a thalamocortical circuit framework to explore mechanisms of movement generation \citep{Logiaco2021}. Dynamical motifs in the activity could be further explored through dynamical systems frameworks, incorporating dimensionality reduction techniques to look at patterns of ‘trajectories’ in neural population activity which may be task-specific \citep{Churchland2012, Shenoy2013}.

In summary, the implementation of a local connectivity scheme in a spiking neural network model has shown that the topology of the network plays a critical role in the resulting cortical dynamics. Our results support theories of structured cortical circuitry and local, patchy connectivity in the generation of spontaneous activity patterns in the motor cortex. Random networks appear to be more regular or synchronous in their behaviour, while the local connectivity model showed more realistic irregularity, particularly in the large pyramidal, output neurons of layer 5. Our model builds on previous work incorporating local connectivity in a more complex laminar structure and cortical circuit, and reproduces firing patterns comparable to those measured \textit{in vivo} in the motor cortex. The output from our model indicates the importance of including physiologically-based local network topology, which resulted in an increased range and irregularity in firing as well as a slight regularisation in the power spectrum, compared to a random network. 
\section{Conclusion} \label{conclusion}
In developing this model, we have aimed to build on previous modelling work and keep parameters as physiologically realistic as possible to replicate the spontaneous firing activity in the motor cortex. To our knowledge, this is the first physiologically based spiking neural-network model to explore the effects of 3D spatially realistic connectivity on spontaneous neuron firing activity in the motor cortex. The pattern of connectivity is shown to play an important role in the generation of irregular firing dynamics and long-tailed firing distributions of cortical neurons, which could have an impact on criticality and information transmission. With the implementation of local connectivity in a structured cortical circuit, this model links neuroscientific theories of structure to functional network dynamics. 

\section*{Code availability \& License}
Code and data available at \url{https://github.com/MunozatABI/MotorCortex}. This work is licensed under a \href{"https://creativecommons.org/licenses/by-nc/4.0/"}{Creative Commons Attribution-NonCommercial 4.0 International License}.

\section*{Acknowledgements}
The authors thank Gonzalo Maso Talou and Gurleen Singh for their valuable discussions and Mark Sagar for initial project funding and administration. 

\section*{Funding}
L.H. was supported by funding from Callaghan Innovation. The funders had no role in the study design or decision to publish.

\section*{Conflict of interest}
The authors declare no conflicts of interests.


\bibliography{main}



\end{document}